\documentclass[12pt,preprint]{aastex} 
\shorttitle{Magnetized Winds Asymptotics}
\shortauthors{Heyvaerts and Norman}

\begin{document}

\title{GLOBAL ASYMPTOTIC SOLUTIONS 
FOR NON-RELATIVISTIC MHD JETS AND WINDS}

\author{ Jean Heyvaerts } 
\affil{Universit\'e Louis Pasteur, Observatoire de Strasbourg 
\altaffilmark{1,4} }
\email{heyvaert@astro.u-strasbg.fr}

\and
 
\author{Colin Norman }
\affil{Department of Physics and Astronomy, Johns Hopkins University \\
and Space Telescope Science Institute\altaffilmark{2,3}  }
\email{norman@stsci.edu}
\bigskip


\altaffiltext{1}{Observatoire, Universit\'e Louis Pasteur, 
11 Rue de l'Universit\'e, 67000 Strasbourg, France            }

\altaffiltext{2}{Space Telescope Science Institute,
3700 San Martin Drive, Baltimore, MD 21218}

\altaffiltext{3}{Department of Physics and Astronomy, Homewood Campus,
The Johns Hopkins University, 3400 North Charles Street, Baltimore, MD
21218}

\altaffiltext{4}{Visiting Scientist at Space Telescope Science Institute
and Department of Physics and Astronomy, Johns Hopkins University}

\begin{abstract}

We present general and global 
analytical solutions, valid from pole to
equator, for the asymptotic structure of non-relativistic, rotating,
stationary, axisymmetric, polytropic, unconfined, perfect MHD
winds. The standard five lagrangian first integrals along field lines
are assumed known.

The asymptotic structure of such winds consists of field-regions
virtually devoid of poloidal current.  We show that an Hamilton-Jacobi
equation, or equivalently a Grad-Shafranov equation, gives the
asymptotic structure in the field regions. These field regions are
bordered by current-carrying boundary layers around the polar axis and
near null magnetic surfaces. Current closure is achieved in a number
of separate cells bordered by null surfaces. The solution is given in
the form of matched asymptotics separately valid outside and inside
these boundary layers. The polar boundary layer is pressure-supported
against the pinching force exerted by the axial poloidal current and
has the structure of a current pinch, while the null-surface boundary
layers have the structure of current sheet pinches. We establish a
consistency relation between the residual poloidal current at large
distances and the axial pressure. We find a similar relation for the
current sheets at null surfaces. We further consider the case where
the polar boundary layer is force free. The geometry of magnetic
surfaces in all parts of the asymptotic domain is explicitly deduced
in terms of the first-integrals.

The solutions have the following general properties:

I. For winds which are kinetic-energy dominated at infinity we derive
WKBJ analytic solutions whose magnetic surfaces focus into
paraboloids.  The current slowly weakens as the inverse of the
logarithm of the distance to the wind source while the axial plasma
density falls-off as a negative power of this logarithm.

II. For winds carrying Poynting flux at large distances the solutions
asymptotically approach to nested cylindrical and conical magnetic
surfaces.

\end{abstract}


\section{Introduction}
\label{intro}

We have previously established \citep{HeyvNorman89}
that any stationary axisymmetric magnetized wind will collimate
at large distances from the source, under perfect MHD conditions and
polytropic thermodynamics, to paraboloids or cylinders along the
symmetry axis according to whether the electric current (or Poynting
flux) brought by the wind to infinity asymptotically vanishes or is
finite.  Our result however did not discriminate between these two
possibilities, nor described any global asymptotic solution. The aim
of this paper is to partially fill this gap.  We consider here
non-relativistic winds and concentrate on describing analytically
their structure in the aymptotic region, for a supposedly given set
of first integrals and for different, {\it{a priori}}
possible, values of the circumpolar current at large distances. 

We show here that the asymptotic structure of rotating MHD winds
consists of vast regions where the poloidal current density is
negligibly small, bounded by thin regions where residual asymptotic
poloidal current flows.  These regions have at large distances the
character of boundary layers. They are located in the vicinity of the
polar axis and of the null magnetic surfaces.  We obtain solutions
valid in all these regions separately and produce a global solution by
asymptotic matching.

The specific topic which we are pursuing here is the
construction of a general, non-self-similar, asymptotic solution
globally valid from the polar axis to the equator, for a given set of
supposedly known first integrals.

There has been in the past few years considerable progress in this
field, both in the derivation of special exact solutions to the wind
equation and in numerical solutions to them, both time dependent and
stationary.  
Focusing of magnetized winds appears to be a robust property of
rotating MHD winds \citep{blandfordpayne,HeyvNorman89}.  
Most analytical solutions involve some sort of self-similarity
\citep{Lovelaceetal91,ContopoulosLovelace94,EOstriker97,
Trussonietal97,VlahakisTsing98,VlahakisTsing99}. \citet{LyndenBell96}
has constructed quasi-static, force-free collimated
structures that arise naturally from a wound up magnetic field pushing
out from a disk. The dynamics of such structures 
has been studied by \citet{Kudohetal2002}. 

\citet{tsinsauty92a,tsinsauty92b}, \citet{sautyetal94} and \citet{sautyetal99} 
have analytically considered
particular models of non-polytropic winds and found that
non-cylindrical asymptotics can be achieved only when magnetic
pinching is negligible and there is over-pressure in the vicinity of
the axis. 

Shu and collaborators have extensively developed a X-wind model for
outflows and in \citet{Shuetal95} (plus references therein) they
actually solved a Grad-Shafranov equation for the shape of the
magnetic surfaces \citep{shafranov}. They have an inner boundary
condition of finite pressure. The launching region for T-Tauri winds
has been examined recently by \citet{anderson}. Interesting disk
instability can be initiated by wind driven angular momentum loss
\citet{cao}. \citet{tomisaka} has found that the collapse of rotating
magnetized molecular clouds and the resulting bipolar outflows are
inextricably related.

Transfield force balance has also been studied by numerical
simulations. These simulations studied hoop stresses collimation in
different parameter regimes. \citet{Ustyugovaetal95} and
\citet{Romanovaetal97} obtained two-dimensional solutions for jets
emitted by a source with Keplerian rotation and confirmed the focusing
effect of the hoop stress, as did
\citet{ouyedpudritz97a,ouyedpudritz97b,ouyedpudritz99}
who also
found time-dependent behaviour. \citet{LeryFrank}
found by
two-dimensional simulation that winds orginating from a disk with a
Keplerian rotation profile have a dense, current-carrying, central
core surrounded by an almost current-free region. \citet{Ustyugovaetal99}
numerically calculated this same problem using a model with a hot wind
source region in the vicinity of the polar axis and found very little
collimation within the simulation region.  \citet{Ustyugovaetal00}
studied formation of collimated Poynting jets associated with an
uncollimated hydromagnetic outflow.  \citet{bogovalovtsin99}
numerically found collimation to be most effective for a particular
class of objects which they describe as "efficient rotators".  In a
subsequent paper \citet{tsinbogovalov00} 
discussed the
efficiency of collimation in the case of the solar wind. Boundary
conditions at the base of the flow were found to be
important. \citet{bogovalov96,bogovalov00} 
studied cold non-stationary flow from a
monopole and found significant Poynting flux conversion to kinetic
energy.

Since \citet{HeyvNorman89},
papers similar in spirit to
our own research have appeared. \citet{Li93} and 
\citet{begli94} 
studied kinetic energy dominated winds and obtained asymptotic
solutions that agreed with the paraboloidal solutions we had
previously obtained. \citet{Tomimatsu95}
constructed solutions in
different regions including the pole and the equatorial null surface
quite similar in spirit to the analysis presented here. Okamoto and
collaborators \citep{okamoto99,okamoto00,beskinokamoto2000}
emphasized the issue of current closure and its effect on the geometry of the
solution. These issues which are, in fact, important only in boundary
layers such as the equatorial current sheet, are now fully treated in
this paper.

Our presentation is structured as follows.  Section (\ref{basics})
reviews the basics of rotating MHD winds.  Section (\ref{secTFasympt})
deals with the field-regions where almost no poloidal current
flows. For Poynting jets the transfield equation becomes an Hamilton--Jacobi equation (Eq.(\ref{Iassgeneral})), or equivalently
a Grad-Shafranov equation (Eq.(\ref{TFassoperatorform})), which we solve
in section (\ref{secasymptfield}) (Eq.(\ref{circleray})).  Kinetic
winds are solved using a WKBJ approximation.  In section
(\ref{secpolarbl}) we obtain the solution for the polar boundary layer
(Eqs.(\ref{parabpsiblofx})--(\ref{parabablofx})). This solution, which
is similar to a Bennet pinch, is then matched to the field-region
solution and a relation between the asymptotic current and the axial
pressure is derived (Eq.(\ref{bennetparab})). The case of
the force-free polar boundary layer is also discussed, giving a relation
similar to the standard Bennet pinch (Eq.(\ref{Bennetffapp})). 
The matching procedure specifies the density along 
the polar axis (Eqs.(\ref{nzeroofRlog})-(\ref{bennetparab})), 
the asymptotic circumpolar current and
the radius of the current-carrying region (Eq.(\ref{radiusofpolarBL})).
A slow logarithmic decline of the axial density and current is found
in case of kinetic-energy-dominated winds, justifying the WKBJ
approach. In section (\ref{nullsurfaceBL}) we similarly obtain a
solution valid in the vicinity of a null magnetic surface
(Eqs.(\ref{alphanullparam})--(\ref{psinullBLofX})) and match to the
field-region solution. The gas pressure at the null surface balances
the toroidal magnetic pressure outside the
sheet (Eq.(\ref{behaverhoequat})).  In section (\ref{sectionshape}) the
shape of the magnetic surfaces has been calculated in all regions and
in both regimes.  In the paraxial field-region of kinetic winds it is
given by Eqs.(\ref{paraboloidshape})--(\ref{cospsiexponent}).  In the
polar boundary layer it is given by Eq.(\ref{shapeincenterofbl}).
Inside an equatorial null-surface boundary layer of a kinetic wind it
is given by Eq.(\ref{zofRinequatBLparab}), while inside the
equatorial null-surface boundary layer of a Poynting jet it is given
by Eq.(\ref{zofRinequatBLcyl}). 
Our conclusions regarding the general
properties of non-relativistic rotating MHD winds are presented in
section (\ref{secconclusion}).

\section{Axisymmetric Stationary MHD Flows}
\label{basics}

\subsection{Notation and Definitions.}
\label{notations}

The formulation of stationary axisymmetric rotating MHD winds have
been presented in a number of papers 
\citep{weberdavis67,okamoto75,mestelbook,sakurai85,HeyvNorman89,
HeyvMiniato}.
Cylindrical coordinates $r$, $\theta$, $z$ are used
with unit vectors $\vec{e}_r$, $\vec{e}_{\theta}$, $\vec{e}_z$.  The
spherical distance of a point to the origin is denoted by $R$. Vectors
are split into toroidal and poloidal parts, indicated by subscripts
$\theta$ and P respectively. The MKSA system of units is used with
$\mu_0$ being the magnetic permeability of free space. The poloidal
magnetic field can be expressed in terms of a magnetic flux function
$a(r,z)$, such that
\begin{equation}
\vec{B}_P = - {1\over r}{\partial a \over \partial z} \  \vec{e}_r
+ {1\over r}{\partial a \over \partial r} \ \vec{e}_z .
\label{Bpversusa} 
\end{equation}
The magnetic flux $\Phi_m$ through a circle
of radius $r$ centered on the axis at altitude $z$ is
$\Phi_m = 2 \pi a(r,z)$.
The magnetic surfaces, generated by 
the rotation of field lines about the axis,
are surfaces of constant value of  $a(r,z)$.
The value of $a$ labels magnetic surfaces
or field lines.

\subsection{First Integrals.}
\label{firstintegrals}

 Mass conservation implies
\begin{equation}
\rho \vec{v}_P = \alpha(a) \vec{B}_P ,
\label{defalpha} \end{equation}
where $\alpha(a)$ is a first integral.  Flux freezing implies that
\begin{equation}
\rho v_{\theta} = \alpha (a) B_{\theta} + \rho r \Omega (a)   ,
\label{defomega} \end{equation}
where $\Omega (a)$ is a second first integral.  Conservation of
specific angular momentum implies,
\begin{equation}
L(a) = r v_{\theta} - {r B_{\theta} \over{ \mu_0 \alpha(a)}}. 
\label{defL} \end{equation}
Conservation of the specific energy $E(a)$ implies:
\begin{equation}
E(a) = {1\over2}v_P^2 +{1\over2}v_{\theta}^2 +{\gamma\over \gamma - 1}
{p\over\rho} +\Phi_{G} - {{r B_{\theta} \Omega(a)} \over{ \mu_0 \alpha(a) }} 
\label{Bern} \end{equation}
$E(a)$ is the total specific energy carried by the wind on magnetic
surface $a$, in kinetic and Poynting flux form.  $\Phi_{G}$ is the
gravitational potential.  Equation (\ref{Bern}) is the Bernoulli
equation.

The thermodynamics is described by a lagrangian polytropic law,
relating pressure and density, of the form
\begin{equation}
p = Q(a) \rho^{\gamma}
\label{polytrop} \end{equation} 
This defines a fifth first integral, $ Q(a)$, the polytropic entropy.
Equation (\ref{polytrop}) may represent 
adiabatic or more complex thermodynamics.

\subsection{Semantics.}
\label{semantics}

We define terms that will be used in this paper.
The {\it asymptotic domain} consist of all points
which are located on their own magnetic surface
far away from the Alfv\'en point.
A  magnetic surface is said to
be {\it asymptotically parabolic} if $r$ and
 $(z/r)$ both become infinite.
This does not imply that $z$
should vary as a power law of $r$ at fixed $a$. A magnetic surface is said
to be {\it asymptotically conical} if $r$ becomes  infinite at
large distances while $(z/r)$ approaches a finite limit
$\tan \psi_{\infty} (a) $. This does not imply
that $(z - r \tan \psi_{\infty})$ should approach a finite limit.
Conical magnetic surfaces may have
parabolic branches. An {\it asymptotically cylindrical} magnetic
surface is one on which the axial distance $r$ approaches a
finite limit $r_{\infty}(a)$. 

When the asymptotic magnetic structure consists of cylindrical
surfaces nested inside conical ones, the cylindrically focused one are
refered to as forming the {\it jet}, and the other surfaces as forming
the {\it{conical wind}}. The {\it jet edge} is defined as being the
magnetic surface which separates asymptotically cylindrical from
asymptotically conical surfaces.  This terminology does not imply that
only cylindrical asymptotics should give rise to structures that would
appear to the observer as an astrophysical jet.

A {\it neutral or null magnetic surface} is one along which the
poloidal field vanishes. The toroidal field then also vanishes on it.
The immediate vicinity of neutral magnetic surfaces is of special
interest, being regions where electric currents flow in the asymptotic
domain. We refer to these regions as {\it neutral surface boundary
layers}. Similarly, the vicinity of the polar axis is a region of
electric current flow. We refer to it as the {\it polar boundary
layer}. Outside the boundary layers is the {\it field-region}.

The space between two neighbouring neutral magnetic surfaces
is refered to as a {\it cell}.
The total electric current through a circle of axis $z$
passing through the point $(r,z)$ is:
\begin{equation}
J(r,z) = 2 \pi r B_{\theta} /\mu_0
\label{physcurrent}
 \end{equation} $J$ vanishes at neutral surfaces and is negative for
positive $\Omega$ and $\alpha$.  Therefore we use the quantity
\begin{equation}
I = - {{r B_{\theta}}\over{\mu_0}}
\label{IofBtheta} \end{equation}
which is positive for positive $\Omega$ and $\alpha$ and proportional
to the current $J$. We refer to it as the current, noting that the true
physical current is $J = - 2 \pi I$.

Strucures which convey a finite circumpolar electric current to
infinity are refered to as {\it Poynting jets}. {\it Kinetic winds}
carry no Poynting flux, or current, at infinity.

\subsection{Bernoulli Equation}
\label{BernandTransfield}

The toroidal components of the velocity and of the magnetic field 
can be obtained from the angular momentum conservation and isorotation 
law in terms of $\rho$ and $r$ as
\begin{equation}
v_{\theta} = {L \over r} + {\rho\over r} \  
{L - r^2 \Omega \over \mu_0 \alpha^2 - \rho},
\label{vtheta} \end{equation}
\begin{equation}
B_{\theta} = {\mu_0 \alpha  \rho \over r} \  {L - r^2 \Omega \over \mu_0 \alpha^2 - \rho} .
\label{Btheta} \end{equation}
For well-behaved solutions the numerators must vanish when the
denominators vanish. The cylindrical radius $r$ equals the Alfv\'en radius,
$r_A(a)$, defined by $\Omega(a) \ r_A^2(a) = L(a)$, when the density
$\rho$ equals the Alfv\'en density, $ \rho_A(a) = \mu_0 \alpha^2(a)$.
The Alfv\'en radius defines the position of the Alfv\'en point on
field line $a$.  Asymptotically, for $ r \gg r_A$, the azimuthal
velocity vanishes while the variable $I = -rB_{\theta}/{\mu_0}$
approaches the value
\begin{equation}
I  \rightarrow  {{\rho r^2 \Omega(a)}\over{\mu_0 \alpha(a) }}
\label{rBthetaass} \end{equation}

\subsection{Transfield Equation}
The projection of the equation of motion on the normal to the magnetic
surfaces gives the transfield equation, or generalized Grad-Shafranov
equation (eq. (\ref{transfield}) of appendix
\ref{appreducTFcurvature}). It determines the shape of magnetic
surfaces in terms of the first integrals and of the density.  The set
of coupled Bernoulli and transfield equations is to be solved.  As
shown in appendix \ref{appreducTFcurvature}, the transfield equation
can be reexpressed such as to give the curvature ($d\psi/ds$) of
poloidal field lines in terms of various surface functions and of the
current variable $I$.  The resulting form of the transfield equation
is:
\begin{eqnarray}
\left(1 - {\rho \over \rho_A} \right) v_P^2{ d \psi \over ds}
= {{1}\over{\rho}} {\vec{\nabla}a \over \vert \nabla a \vert }
\cdot \vec{\nabla} 
\left( {{ \vert \vec{\nabla} a \vert^2 }\over{ 2  \mu_0 r^2}} 
+ Q \rho^{\gamma} \right)
+{\vec{\nabla}a \over \vert \nabla a \vert } \cdot \vec{\nabla} \Phi_{G}
\nonumber \\
- \left( L - I/\alpha\right)^2 {1 \over r^3}
{ \partial a/ \partial r \over \vert \nabla a \vert }
+{1 \over \rho r^2}
{\vec{\nabla}a \over \vert \nabla a \vert } \cdot \vec{\nabla}
\left({\mu_0 I^2 \over 2}\right)
\label{TFcurvform} \end{eqnarray}
The terms on the right are the components of the poloidal magnetic
pressure gradient, gas pressure gradient, gravity, centrifugal force
and hoop stress respectively perpendicular to magnetic surfaces.  The
hoop stress is proportional to $\vec{j}_P \times
\vec{B}_{\theta}$ and therefore vanishes with $I$.

\section{Field Regions  }
\label{secTFasympt}

\subsection{Asymptotically Dominant Forces.}
\label{dominantforces}

In the large-$z$ limit, Eq.(\ref{TFcurvform}) simplifies. The poloidal
velocity $\vert v_P \vert$ reaches a terminal value and the curvature
term on the left of equation (\ref{TFcurvform}) approaches zero. A
definite ordering between the terms on the right of equation
(\ref{TFcurvform}) occurs on magnetic surfaces on which both $z$ and
$r$ tend to infinity. On asymptotically cylindrical magnetic surfaces,
$r$ approaches a finite limit $r_{\infty}(a)$.  However, when
$r_{\infty}(a) \gg r_A(a)$ it still remains possible to use the
large-$r$ approximation. 
In the large $r$, large $z$ and small $(\rho / \rho_A)$ limit,
noting that $\rho r^2$ is bounded from above \citep{HeyvNorman89} and that
the gravity term declines as $1/r^2$ while the centrifugal force term declines
as $1/r^3$, Eq.(\ref{TFcurvform}) reduces to:
\begin{eqnarray}
v_P^2{ d \psi \over ds}
={r^2 \Omega \over \mu_0 \alpha I} {\vec{\nabla}a \over \vert \nabla a \vert }\cdot
\vec{\nabla} \left({{\nabla a}^2 \over 2  \mu_0 r^2} + Q \rho^{\gamma} \right)
+ {\Omega \over \alpha} {\vec{\nabla}a \over \vert \nabla a \vert }\cdot \vec{\nabla} I
\label{TFcurvass} \end{eqnarray}

When $I$ approaches a finite value, the hoop stress term,
proportional to $\vec{n} \cdot \vec{\nabla} I$ decreases on conical
magnetic surfaces as $(1/r)$ since $\vec{n} \cdot \vec{\nabla} I
\approx I/r$.  The hoop stress force then dominates all the other
terms on the right of (\ref{TFcurvass}). The poloidal field $B_P$
decreases in this case as $(1/r^2)$.  The gradient of the poloidal
magnetic pressure then drops with axial distance as $(1/r^5)$, so that
the corresponding term in Eq.(\ref{TFcurvass}) declines as $(1/r^3)$.
According to Eq.(\ref{defalpha}), the density decreases as $B_P$ when
the velocity has reached its terminal value so that the gas pressure
gradient scales as $(1/r^{2 \gamma +1})$ and the pressure term in
Eq.(\ref{TFcurvass}) declines as $(1/r^{2 \gamma - 1})$.  This decline
is slower than that of the poloidal magnetic pressure if the
polytropic exponent $\gamma$ is less than 2, which we assume. The
poloidal magnetic pressure is negligible compared to the gas pressure
except in the ultra-cold limit where $Q$ vanishes. Therefore, we
retain the gas pressure on the right of Eq.(\ref{TFcurvass}). The case
of cold winds in which the poloidal magnetic pressure is retained is
discussed in section (\ref{appforcefreepolarbl}).

Kinetic winds are characterized by $r \vert \vec{\nabla} a \vert $ and
 $\rho r^2$ approaching zero.  The hoop stress scales as $\nabla I^2
 \sim \rho^2 r^3$.  The gas pressure gradient term scales as
 $\rho^{\gamma}/r$, and the poloidal magnetic pressure force $ \sim
 \rho^2 /r$.  Again, gas pressure dominates over poloidal magnetic
 pressure, except for very cold winds. The hoop stress term dominates
 over the gas pressure force if $ \rho^2 r^4 \gg \rho^{\gamma}$. This is so in
 the kinetic wind solutions derived below, since $\rho r^2$ is found
 to decline only as an inverse power of $(\ln r)$.

\subsection{Transfield Equation}
\label{subsecTFinfield} 

Neglecting pressure, the transfield equation (\ref{TFcurvass}) then further reduces to:
\begin{equation}
 v_P^2 { d \psi \over ds} =
+{\Omega \over \alpha}  {\vec{\nabla}a \over \vert \nabla a \vert }.
\vec{\nabla} I
\label{TFasswithoutP} \end{equation}
Eq.(\ref{TFasswithoutP}) is equivalent to Okamoto's conclusion
that $j_{\parallel}B_t = \rho v_p^2 /R_c$, $R_c$ being the curvature
radius of poloidal field lines \citep{okamoto99}.  
Note however that the centrifugal
force associated with the poloidal field line curvature (on the left
of equation (\ref{TFasswithoutP})) is also negligible.  Since the
poloidal current is bounded from above, the right of equation
(\ref{TFasswithoutP}) could {\it a priori} scale as $1/r$, but the
curvature $(d \psi/ds)$ must decline with $r$ faster than
$1/r$. Otherwise the poloidal field lines would not be well-behaved
\citep{HeyvNorman89}.  The left of Eq.(\ref{TFasswithoutP})
is therefore of order $(v_P^2/R_c)$ where $R_c$ is the 
curvature radius, much larger than $r$. To lowest order in
the small parameter $(r/R_c)$ Eq.(\ref{TFasswithoutP}) therefore reduces to the
vanishing of its right-hand-side and becomes:
\begin{equation}
\vec{\nabla}a \cdot
\vec{\nabla} I =  0
\label{solvability} \end{equation}
This statement has been presented in \citet{HeyvNorman89} 
as a solvability condition
at infinity. It implies that almost no poloidal electric current
density is present in these regions.

\subsection{Existence of Current Carrying Boundary Layers and Electric Circuit}
\label{subsecexistBL}

Any physical current system must be closed. Current closure causes $I$
to depend on $a$ wherever current is flowing. This implies that some
terms of Eq.(\ref{TFcurvass}) should balance hoop stress where
current flows. Since this is not possible wherever $\nabla \approx
1/r$, this implies that regions of current flow must take the form of
boundary layers in which gradients are large. Note that the hoop
stress vanishes where the current vanishes. Boundary layers must then occur in
the vicinity of the polar axis, where $B_{\theta}$ vanishes for finite
current density at the axis, and near neutral surfaces, where the hoop
stress vanishes with $B_{\theta}$. Indeed, the toroidal field
component is generated from the poloidal one by the rotation, but the
poloidal field vanishes at a neutral surface. In the case of a dipolar
symmetry the equatorial plane is a neutral surface.

We then conclude that Eq.(\ref{solvability}) applies
in the field-region of the asymptotic domain, except in the boundary layers.
Eq.(\ref{solvability}) integrates along any trajectory
 orthogonal to magnetic surfaces as long as no null surface (in the
 vicinity of which equation (\ref{solvability}) ceases to be valid) is
met. The result is:
\begin{equation}
I = I_{\infty}(b)
\label{Iofb} \end{equation}
where $b$ is a label for an orthogonal trajectory. To be specific,
we define it as being the value of the $z$-coordinate on this
orthogonal trajectory at the polar axis.  An extra label, which we
omit, should precise on which segment of the orthogonal trajectory
Eq.(\ref{Iofb}) applies.  Note that the magnetic surface label, $a$,
and the orthogonal trajectory label, $b$, could constitute a set of
orthogonal coordinates in the poloidal plane that could be substituted
to $r$ and $z$, with $a$ playing the role of an angular variable and
$b$ the role of a radial distance variable.

Eq.(\ref{solvability}) states that the component of the poloidal
current density parallel to $\vec{B}_P$ vanishes. Since
$I_{\infty}(b)$ approaches a constant (possibly zero), so does also
the other component of $\vec{j}_P$. Thus, the hoop stress $\vec{j}_P
\times \vec{B}_{\theta}$, approaches zero, as does $\vec{j}_{\theta}
\times \vec{B}_P$, since $\vert \vec{B}_P \vert$ declines rapidly with
distance. Hence, the Lorentz force $\vec{j} \times \vec{B}$ approaches
zero.  This does not imply that the field becomes force free, because
this decline is due to both current and field approaching zero. It is
in fact quite interesting to analyze how the different components of
the Lorentz force individually approach zero at large distance. This
requires, however, a more precise knowledge of the solution than we
have at this point. Therefore this discussion is
postponed to appendix {\ref{appcomponentsofLorentz}}. We emphasize
that the asymptotic regime discussed in this paper is not force-free.

\begin{figure}
\epsscale{0.75}\plotone{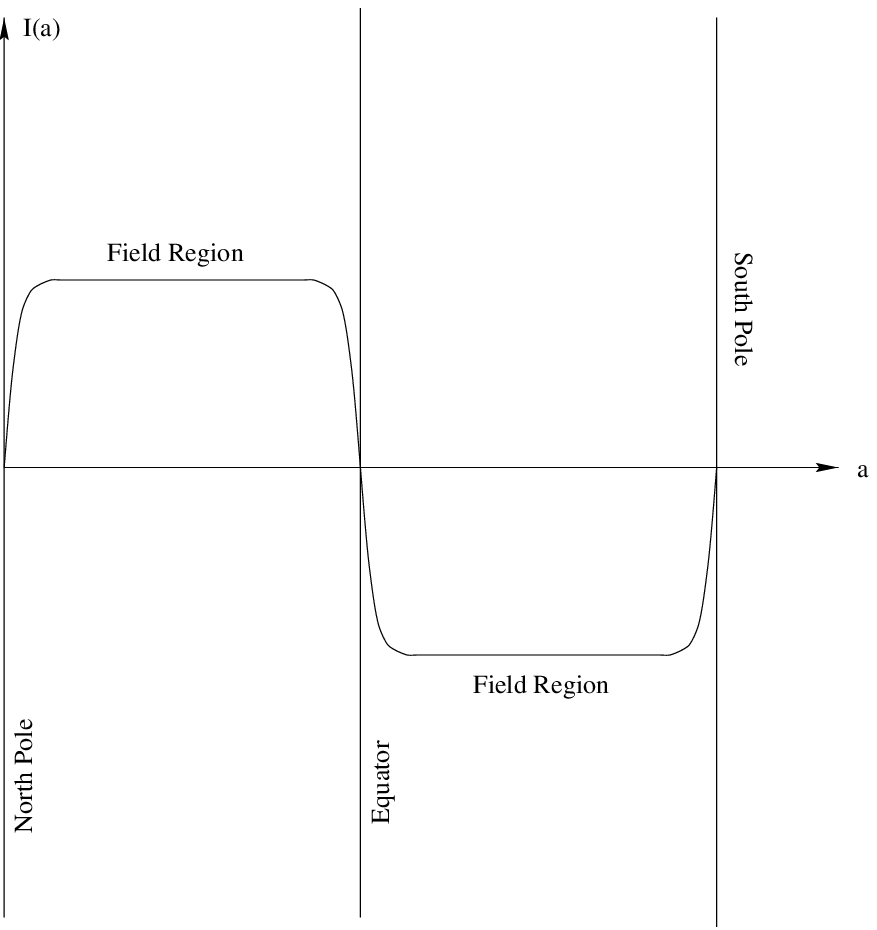}
\caption{Variation of the integrated current in the asymptotic
domain following an orthogonal trajectory
from the north polar axis to the south polar axis 
versus the flux variable. The
poloidal field is assumed to have dipolar symmetry.
\label{fig1}}
\end{figure}

\begin{figure}
\plotone{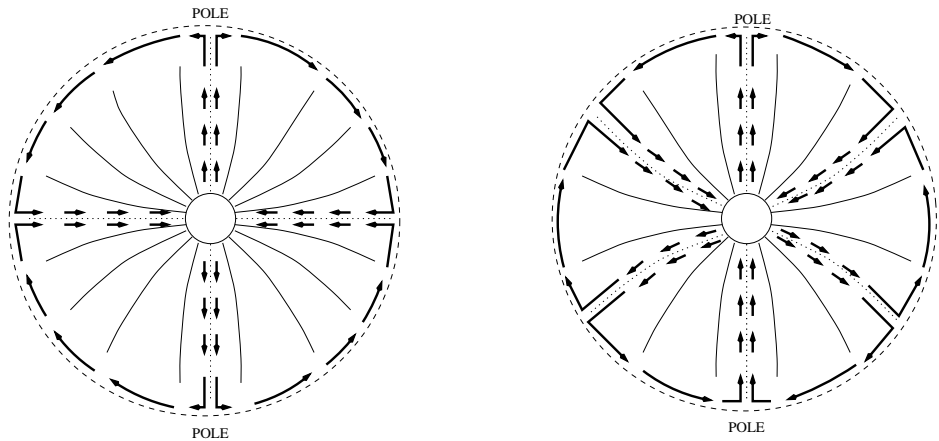}
\caption{A schematic representation of the poloidal current circuit in the 
asymptotic domain, in the case of a field with two different polarity zones
(dipolar symmetry, left) on the wind source (central object), and in the case
of three different polarity zones (quadrupolar symmetry, right) on the wind
source. Current is concentrated in boundary layers near the pole and the
neutral magnetic surfaces (dotted lines). The heavy arrows indicate the main
poloidal electric current channels. The dashed line is meant to represent a
surface at infinity. In the case of kinetic winds there is also a very weak
and diffuse current in the field regions between the regions of largest 
current flow so that the boundary layer current vanishes at infinity.
\label{fig2}}
\end{figure}

According to Eq.(\ref{Iofb}), the enclosed poloidal current is
approximately constant through any circuit drawn on surfaces
orthogonal to magnetic surfaces and running between two successive
boundary layers.  The current enclosed in a circle of increasing
radius drawn on one such surface would increase from zero on the polar
axis to some value at the outer edge of a polar boundary layer. The
current remains almost constant up to the next null-surface, at which
it returns to zero (Fig \ref{fig1}) through a boundary layer, vanishing at
the null surface.  Needless to say,
Eqs.(\ref{solvability}) and (\ref{Iofb}) are only approximate results. 
We do not mean that the current density is strictly zero
in field regions, but only that it is so small that the total electric current
circulating between two successive boundary layers 
through a surface of constant-$b$ is much less than the current in
the boundary layers themselves. Nevertheless a small current in field regions
must be present for $I_{\infty}$ to be a (slowly varying) function of $b$.
This picture generalizes for any number of null
surfaces and the current system consists of more than just two
cells (Fig \ref{fig2}), in each of which the electric circuit separately
closes.  $I_{\infty}(b)$ approaching zero as $b \rightarrow
\infty$ means that the poloidal current circuit closes in each cell at
a finite distance.  By contrast, $I_{\infty}(b) $ approaching a
finite value, $I_{\infty}$, implies that this circuit would close at
infinity. In reality, the wind has not been blowing for an
infinite time and it is externally bordered by a time-dependent
region and an expanding shock system. This region plays the role
of the "region at infinity" in the present stationary model: this
is where the residual 
electric currents flow from the circumpolar region
to the return-current boundary layers shown in Fig.(\ref{fig2}).

Regions about the polar axis and the null surfaces must have the
character of thin boundary layers since in the asymptotic domain, the
gas pressure is a subdominant force that can balance Lorentz forces
only in a small region about those special places where the latter
vanishes. These boundary layer sheets must be thin and there must be
toroidal magnetic pressure equilibrium accross them. The total
poloidal current then only changes sign at their crossing as
$\vec{B_{\theta}}$ reverses. 

Our aim now is to solve the wind equations in the asymptotic domain,
both in field-regions and in current-carrying boundary layers and
obtain the resulting shape of the magnetic surfaces.

\subsection{Asymptotic Grad-Shafranov Equation}

From Eq.(\ref{TFasswithoutP}), it is possible to restate
Eq.(\ref{solvability}) as an equation similar to the familiar
Grad-Shafranoff equation of magnetohydrostatics. Using the identity
(\ref{rotvpcrossvp}) of appendix \ref{appreducTFcurvature} and
Eqs.(\ref{rBthetaass}), (\ref{Bpversusa}) and (\ref{defalpha}), the
equation (\ref{TFasswithoutP}) becomes:
\begin{equation}
r {{d \psi}\over{ds}} = {{\mu_0 I}\over{\Omega v_{\infty} }}
\left( {{r^2 \Omega}\over{\mu_0 I}} \  {\mathrm {div}} 
\left( {{\Omega \vec{\nabla}a}\over{ \mu_0 I}} \right) 
- {{\partial}\over{\partial a}} \left( {{v_{\infty}^2}\over{2}} \right)
\right)
\label{transfotoDel1}
\end{equation}

The above discussion indicates that the curvature radius 
should be much larger than r, so that the left of Eq.(\ref{transfotoDel1}) 
is negligible. Expanding the divergence term and using Eqs.(\ref{defalpha}),
(\ref{Bpversusa}) and (\ref{rBthetaass}) to express $r \vert \vec{\nabla} a \vert$,
Eq.(\ref{transfotoDel1}) is eventually brought to the form of a
quasi-linear elliptic equation:
\begin{equation}
r^2 \Delta a = {{\partial}\over{\partial a}} \left( 
{{ \mu_0^2 I^2 v_{\infty}^2}\over{ 2 \Omega^2}} \right)
\label{TFassoperatorform}
\end{equation}
The boundary conditions to Eq.(\ref{TFassoperatorform}) are that
$a =0$ along the polar axis and that $a =A$ on the equatorial plane. These
are consistent, when $I$ is constant and non-zero,
with $a$ depending only on the latitude angle $\psi$.
In this case Eq.(\ref{TFassoperatorform})
becomes an ordinary differential equation
for $a(\psi)$, which reduces to the form
of Eq.(\ref{eqpsiofageneral}). 
Eq.(\ref{TFassoperatorform}) is equivalent to the other forms obtained above, 
in particular it implies  Eq.(\ref{solvability}). 
It is also equivalent to the 
alternative forms derived below, in particular Eq.(\ref{Iassgeneral}).
Using Eqs.(\ref{rBthetaass}), (\ref{Bpversusa}) and (\ref{defalpha}),
an equivalent form of Eq.(\ref{TFassoperatorform}) is in fact found to be:
\begin{equation}
\Delta a  = \vec{\nabla}a \cdot \vec{\nabla} 
\left( \ln \left( r \vert \vec{\nabla}a \vert \right) \right) 
\label{TFassgeometricalform}
\end{equation}
which can be brought by denoting the normal unit vector to poloidal field 
lines by $\vec{n} = - \vec{\nabla}a / \vert \vec{\nabla}a \vert$,
to the form
\begin{equation}
{\mathrm{div}} \left({{\vec{n}}\over{r}} \right) = 0
\label{divnoverr}
\end{equation}

\section{Solutions in Field Regions}
\label{secasymptfield}

\subsection{An Hamilton-Jacobi Equation }
\label{subsecHamJacob}

Noting that the velocity on each surface
reaches a terminal velocity $v_{\infty}(a)$,
Eqs. (\ref{defalpha}), (\ref{Bern}) and
(\ref{Btheta}) reduce in the asymptotic limit to:
\begin{equation}
\rho r  v_{\infty} (a) =  \alpha \vert \vec{\nabla} a \vert
\label{defalphaasympt}
\end{equation}
\begin{equation}
I =  + {{\rho r^2 \Omega}\over{\mu_0 \alpha}}
\label{Iasympt}
\end{equation}
\begin{equation}
E = {{v_{\infty}^2}\over{2}} + {{I \Omega}\over{\alpha}}
\label{Bernasympt}
\end{equation}
Eliminating $v_{\infty}$ and $\rho$ between
Eqs.(\ref{defalphaasympt}), (\ref{Iasympt}) and (\ref{Bernasympt}), an
expression of $r \vert \vec{\nabla} a \vert$ is obtained:
\begin{equation}
r \vert \vec{\nabla} a \vert = {{\mu_0 I \sqrt{2} 
\sqrt{ \left(E - {{I \Omega}\over{\alpha}}\right) } 
}\over{\Omega}}
\label{rgradaofIgeneral}
\end{equation}
This relation can be integrated following  
orthogonal trajectories to
magnetic surfaces. The 
curvilinear abcissa along them is
denoted $\sigma$ and conventionally increases from pole to
equator, so that  $\vert \vec{\nabla} a \vert = da/d\sigma$.
Eq.( \ref{rgradaofIgeneral}) becomes
\begin{equation}
{{d \sigma}\over{r}} =  -
{{\Omega(a) da}\over{\mu_0 I(a,b) \sqrt{2}
\sqrt{E(a) - I(a,b) \Omega(a)/\alpha(a) }   }}
\label{fluxversuscurvilinear}
\end{equation}
In field-regions, Eq. (\ref{fluxversuscurvilinear}) simplifies, since
$I(a,b)$ becomes a function $I_{\infty}(b)$ independent of $a$.
It can be restated as:
\begin{equation}
r \vert \nabla a \vert =  {{ \mu_0 I_{\infty}(b) \sqrt{2}
\sqrt{E(a) - I_{\infty}(b)  \Omega (a) /\alpha (a)}
}\over{
\Omega(a)
}}
\label{Iassgeneral} \end{equation}
If $I_{\infty}(b)$ were to approach a non-vanishing value
$I_{\infty}$ at large distances from the wind source, this equation
would explicitly give the flux distribution in space in the asymptotic
domain.  Particular versions of Eq.(\ref{Iassgeneral}) for cylindrical
or conical magnetic surfaces have in fact been obtained and solved in
our earlier work \citep{HeyvNorman89}.  
Eq. (\ref{Iassgeneral}) improves on this by
not being restricted to either one of these specific geometries. It
implies no {\it a priori} constraints on the structure of the solution
and keeps a fully 2-dimensional character.  In the asymptotic
field-region, far from the neutral surfaces and from the polar axis,
the transfield equation reduces to Eq.(\ref{Iassgeneral}) with
constant $I_{\infty}(b)$, which is of the Hamilton-Jacobi type:
\begin{equation}
r  \vert \nabla a \vert  =  f(a),
\label{HamJacoba} \end{equation}
By defining  
$ S(a)= \int_0^a {{ da'}/{f(a')}} $, 
it can be converted into
\begin{equation}
 \vert \nabla S \vert  =  { 1 \over r},
\label{HamJacobS} \end{equation}

This equation can be restated as $\vec{\nabla}S = - \vec{n} / r$,
implying that $\vec{\nabla} \times (\vec{n} / r) = 0$. This, with
Eq.(\ref{divnoverr}), indicates that $S$ should be an
harmonic function. The solution for $S$ represented by
Eq.(\ref{eqpsiofageneral}) is harmonic.

If the poloidal 
electric current $I_{\infty}(b)$  declines towards zero
at large distances,
the function on the right of
Eq.(\ref{HamJacoba}) also depends on $b$. 
Eq.(\ref{Iassgeneral})
then does not provide the value of the modulus $\vert \nabla a \vert$
as a function of  $a$ and $r$ (as does Eq.(\ref{HamJacoba})) 
because it is not known how $I_{\infty}(b)$ 
declines with $b$.
This difficulty can however be circumvented 
if this variation is so slow that it can be 
treated, as we do below, by a WKBJ type of approach.
Note that the case when
$I_{\infty}(b)$ approaches a non-zero limit,  can be 
dealt with by a WKBJ procedure as well, 
since $I_{\infty}(b)$ is not strictly constant,
but slowly evolves towards its limit. 
Important features of the asymptotic structure,
in particular the difficult question of the connexion between 
the asymptotically cylindrical and the asymptotically conical regions, 
must be approached by considering the 
non-constancy of $I_{\infty}(b)$ as it approaches its limit.

Particular solutions to the Hamilton-Jacobi equation (\ref{HamJacoba})
for constant $I_{\infty}(b)$ can be found by the method of
separation of variables. These are described in appendix
\ref{apppartsolHamJacob}, although the boundary conditions associated
to some of them are different from those we are facing.  These
solutions however exemplify a number of typical structures to which
Eq.(\ref{HamJacoba}) may give rise. General solutions associated with
the boundary conditions relevant to the case of astrophysical MHD
winds can be found by reduction of Eq.(\ref{HamJacobS}) to a
ray-tracing problem.

\subsection{Ray-Tracing}
\label{subsecraytracing}

Equation (\ref{HamJacobS}) is of the form $\vert \nabla S \vert =
N(\vec{r})$, which is the Eikonal equation for the propagation of
waves in a medium of index of refraction $N(\vec{r})$. The wave fronts
are represented by surfaces of constant $S(\vec{r})$.  They may be
found by ray-tracing methods, the rays being the orthogonal
trajectories to the surfaces.  So, finding the general solution of
Eq.(\ref{HamJacobS}) is equivalent to solving Snell's refraction
equation.  We show in appendix (\ref{appsnell}) the equivalence of the
Hamilton-Jacobi equation with Snell's equation.  The polar and the
equatorial lines, being field lines, must be lines of constant
$S$. The boundary condition to Eq.(\ref{HamJacobS}) is that rays start
perpendicular to the pole and the equator. In the vicinity of a null
surface (e.g. the equator), the simplified asymptotic transfield
equation (\ref{solvability}) is invalid.  We show in section
(\ref{subsecexit}) and in appendix \ref{appangleofexitfromeqbl} that
the angle of the magnetic field lines to the equator at the outskirts
of this boundary layer decreases to zero with increasing $b$,
justifying the view that rays from the field-region must nevertheless
end perpendicular to it.  Similar considerations apply to the vicinity
of the pole and to other null surfaces: rays must also cross them
perpendicularly on both sides.

This optical analogy can be used to find the general solution
Eq.(\ref{HamJacobS}) by writing the appropriate form of Snell's law.
The gradient of the refraction index is in this case perpendicular to
the polar axis. Let $i$ be the angle of incidence between this radial
direction and the tangent to the ray. The ray-tracing equations can be
written as:
\begin{equation}
dr = - d\sigma \  \cos i 
  \qquad \qquad dz = d\sigma \  \sin i  \qquad \qquad {{\sin i}\over{r}} = k
\label{Snell} \end{equation}
where $k$ is a constant. Unlike the boundary condition at the equator,
the condition that $i$ approaches zero at the polar axis is not
restrictive because $N(r)$, being equal to $1/r$, diverges there.
Equations (\ref{Snell}) can be integrated to:
\begin{equation}
(z-h)^2 + r^2 = {{1}\over{k^2}}
\label{circleray} \end{equation}
where $h$ and $k$ are two integration constants.  Eq.(\ref{circleray})
represents a circle centered at $z=h$ on the axis, with a radius $D=
1/k$.  Lines of constant $S$ associated with the solution of equation
(\ref{HamJacobS}) are orthogonal trajectories to a collection of
circles centered on the polar axis.  The boundary condition that the
rays connect perpendicularly to the equator implies that $(h/D)$ should
asymptotically vanish. It is important to note that this is not an
exact, but only an approximate, asymptotic result.  This is consistent
with our earlier results \citep{HeyvNorman89}. We show in appendix 
{\ref{appnoparaboloidswithcurrent}} that other conceivable current-enclosing
geometries are not consistent with a 
source subtending a finite flux.
We have assumed above that a unique
cell extends from pole to equator. The extension to winds with a
larger number of null surfaces is described in appendix
\ref{appcells}.

\subsection{WKBJ Approximation}
\label{WKBapproach}

For $I_{\infty}(b)$ approaching a non-zero value as $b$ grows, the
field-region of the flow consists of asymptotically conical magnetic
surfaces in which current-carrying cylindrical ones are nested.  When
$I_{\infty}(b)$ approaches zero, we shall assume that it does so only
very slowly, so that a WKBJ approach, which considers $I_{\infty}(b)$
as almost constant over large domains of $b$ values, will be possible.
This approach assumes that the flux distribution on an orthogonal
trajectory $b$ changes only very slowly when $b$ increases.  The same
WKBJ treatment can be used when $I_{\infty}(b)$ approaches a non-zero
value, since it does so by only slowly varying. As explained before,
considering this variation is a useful refinement. It is consistent to
WKBJ analyze Eq.(\ref{solvability}) without taking into account any of
the other terms present in Eq.(\ref{TFcurvass}).  The gas and poloidal
magnetic pressure terms have been shown to be negligible to a certain
order in $r^{-1}$.  The relative order of magnitude of the inertia
force associated to the curvature of the poloidal motion will define
the order to which the WKBJ solution is consistent a posteriori.  It
will be shown that this term is indeed insignificant, even when
compared to the gas pressure term, which is itself small in the
field-region.

\subsection{Solution in Field Regions}
\label{subsecfluxdistrgeneral}

Orthogonal trajectories to magnetic surfaces are approximately circles
centered at the origin.  The distribution of flux on such a circle of
radius $b$ can be represented by the distribution of latitude $\psi$
with flux $a$. In the WKBJ approximation, this distribution slowly
changes from one circle to the next, so that $\psi$ depends not only
on $a$ but also weakly on $b$, which we may now identify with 
the spherical distance $R$ since the orthogonal trajectory's label 
in fact coincides with the
distance to the origin all along it. Therefore, magnetic
surfaces may be locally approximated by cones of semi-opening angle
$(\pi/2 - \psi(a,b))$. The equation for these cones is:
\begin{equation}
z = r \tan(\psi(a, b))
\label{eqofcones}
\end{equation}
From this we obtain by differentiation and ignoring the WKBJ dependance
on $b$:
\begin{equation}
\vert \nabla a \vert = - {cos \psi \over r \ \partial \psi /\partial a }
\label{gradaversuspsi}
\end{equation}
From Eq.(\ref{fluxversuscurvilinear}) it results that $\psi(a)$
satisfies the differential equation:
\begin{equation}
{{1}\over{\cos \psi}} \ {{d \psi}\over{d a}} =
{{\Omega(a) }\over{\mu_0 I(a,b) \sqrt{2}
\sqrt{E(a) - I(a,b) \Omega(a)/\alpha(a)}  }}
\label{eqpsiofageneral}\end{equation}
For this relation to give the flux distribution, the variation with $a$ of
$I(a,b)$ must be known, which is the case only in field-regions of
the asymptotic domain, where $I(a,b)$ becomes independent of $a$. In a
current-carrying boundary layer $I(a,b)$ locally depends strongly on
$a$ and should be determined by an analysis of the transfield
equilibrium.  In a field-region, where $I(a,b)$ reduces simply to
$I_{\infty}(b)$, the solution of Eq.(\ref{eqpsiofageneral}) is:
\begin{equation}
\tan (\psi (a,b)) = \tan(\psi (a_1, b))
+ \sinh \left( \int_a^{a_1}
{{1}\over{\sqrt{2} \mu_0 I_{\infty}(b) }} \
{{\Omega(a') da'}\over{
\sqrt{ E(a') - I_{\infty}(b) \Omega(a') / \alpha(a') }
}} \right)
\label{solpsiofa}
\end{equation}
where $a_1$ is a reference flux in the cell under consideration.
If the cell begins at the equator, $a_1$ is the flux variable, $A$,
for the equatorial surface and $\tan(\psi(A,b)) = 0$. 

\subsection{Flux Distribution in Cylindrical Regions of the Field.}
\label{subsecfluxdistrcylindrical}

In a region of the free field where the distribution of flux 
is cylindrical, orthogonal trajectories are better represented as 
straight lines perpendicular to the axis and Eq.(\ref{fluxversuscurvilinear})
integrates to:
\begin{equation}
r_{\infty} (a) = r_{\infty}(a_1) \exp \left( \int_{a_1}^a
{{1}\over{\sqrt{2} \mu_0 I_{\infty} }} \
{{\Omega(a') da'}\over{
\sqrt{ E(a') - I_{\infty} \Omega(a') / \alpha(a') }
}} \right)
\label{rinftyofa}
\end{equation}
where $a_1$ is a reference flux in the cylindrical field-region.
The flux distribution described by Eqs.(\ref{solpsiofa})-(\ref{rinftyofa})
with non-zero $I_{\infty}$ is represented, for arbitrarily chosen functions
$\alpha$, $E$ and $\Omega$, in Fig.(\ref{fig6}). A necessary, but not a
sufficient, condition for such a solution to be 
obtained at very large distances from the source is that the function
$(\alpha E/\Omega)$ has an absolute minimum. 
Whether a particular system, such as an accretion disk 
launching a centrifugal wind, can indeed meet this 
necessary condition can only be decided 
by solving the regularity conditions that this flow 
should satisfy. This point is addressed in the accompanying paper 
\citep{HNIII}, where we discuss whether the condition 
that $(\alpha E/\Omega)$ has a minimum value is sufficient 
to induce cylindrical collimation at
infinity. We show that, in a purely mathematical sense, 
it is not but that in a physical sense 
it should be met in jets of a finite, albeit long, extent. 

\section{The Polar Boundary Layer}
\label{secpolarbl}

\subsection{Solution in the Polar Boundary Layer.}
\label{solutionpolarBL}

The field-region solution does not apply to boundary layers. Within
these boundary layers, the Lorentz force almost vanishes and forces that
would elsewhere be negligible should be locally taken into account.
The discussion of section (\ref{dominantforces}) has shown that the
pressure needs to be considered.  In the case of pressureless winds,
the poloidal magnetic pressure would be the dominant extra force.
This situation is considered for completeness in section
(\ref{appforcefreepolarbl}).  Gravity declines to zero at large
distances and the azimuthal velocity becomes very small at large
$r/r_A$.  Only if the ratio $r/r_A$ does not become very large would
the centrifugal force play a role in the transfield equilibrium.  In
this section, we retain only gas pressure and hoop stress in our
discussion of transfield equilibrium.

Physically, the polar boundary layer then locally has the structure of
a column pinch.  The asymptotic transfield equation can be written as
Eq.(\ref{TFcurvass}), modified by using Eqs.(\ref{defalpha}) 
and (\ref{Iasympt}), which first gives
\begin{equation}
v_{\infty}^2 {{d \psi}\over{ds}} =
{{\Omega}\over{\alpha}} {{\vec{\nabla}a. \vec{\nabla}}\over{ 
\vert \nabla a \vert }}
\left(  {{ \rho r^2 \Omega}\over{\mu_0 \alpha}}
\right)
+ {{1}\over{\rho}} \ {{\vec{\nabla}a. \vec{\nabla}}\over{\vert \nabla a \vert }}
\left(Q \rho^{\gamma}\right)
\label{parabassTF} \end{equation}
We find below that the axial density drops down
with distance only very slowly, so that the
poloidal curvature inertia term on the l.h.s. can also
be neglected, being smaller than
both the hoop stress and the pressure term.
The transfield equation simplifies to:
\begin{equation}
{\Omega \over \alpha}\
\vec{\nabla} a. \vec{\nabla}
\left(  {{\rho r^2 \Omega}\over{\mu_0 \alpha}}
\right)
+ {{1}\over{\rho}} \  \vec{\nabla} a \cdot \vec{\nabla} 
\left(Q \rho^{\gamma}\right) = 0
\label{parabpinchass} 
\end{equation}

Eq.(\ref{parabpinchass}) is readily integrated when
the region of non-negligible pressure 
encompasses little enough flux that
the first integrals $E$, $Q$, $\alpha$ and $\Omega$ 
can be treated as being constant in it, with values
$E_0$, $Q_0$, $\alpha_0$, $\Omega_0$ say. 
In appendix \ref{appfluxinbl}, we
show this to be so
when the asymptotic Poynting flux
is small compared to the kinetic energy flux.
If it is not, the pressure-supported
region remains similar to a column-pinch,
but its structure would now have to be calculated numerically.
Assuming locally constant first integrals, then,
equation (\ref{parabpinchass}) becomes
\begin{equation}
\vec{\nabla}a. \vec{\nabla}
\left(
{{ \rho r^2 \Omega_0^2}\over{\mu_0 \alpha_0^2 }}
+ {{\gamma}\over{\gamma-1}}\ Q_0\rho^{\gamma-1}
\right) = 0
\label{eqparabpinchaxis} \end{equation}
which integrates, following an orthogonal trajectory to
magnetic surfaces, into:
\begin{equation}
{{ \rho r^2 \Omega_0^2}\over{\mu_0 \alpha_0^2 }}
+ {{\gamma}\over{\gamma-1}}\ Q_0\rho^{\gamma-1}
=
 {{\gamma}\over{\gamma-1}}\ Q_0 \rho_0^{\gamma - 1}(b)
\label{solparabpinchaxis} \end{equation}
where $b$ is the radius of this quasi-circular orthogonal trajectory
and $\rho_0(b)$ is the axial density at the distance $b$ from the source.
Equation (\ref{solparabpinchaxis}) can be solved for $r$
in terms of the parameter
\begin{equation}
 x = {{\rho}/{\rho_0}} \label{defdensratio}
\end{equation}
At the polar axis $x = 1$ and far from it 
$x$ decreases to very mall values. Since
$r = b \cos \psi$, this solution for $r(x)$ gives $\psi(x)$
at a given $b$, as expressed by Eq.(\ref{parabpsiblofx}). 
Close to the polar axis, the Bernoulli equation (\ref{Bern}) 
reduces in the asymptotic domain to:
\begin{equation}
{{v_P^2}\over{2}} = 
E_0 - {{\gamma}\over{\gamma -1}} Q \rho_0^{\gamma -1} 
\label{Bernfarnearaxis}
\end{equation}
The Poynting flux is negligible because $r B_{\theta}$ 
vanishes proportionally to $r^2$. On the other hand, 
$\vert \vec{v}_P \vert$ is related to
$\vert \vec{\nabla} a \vert$ by Eqs.(\ref{Bpversusa}) and (\ref{defalpha}).
With the approximation of a constant $\alpha(a)$ in the 
polar boundary layer, this gives, using Eq.(\ref{gradaversuspsi}):
\begin{equation}
\vert \vec{v}_P \vert = {{\alpha_0 }\over{ b^2 \rho_0(b) }}
{{ da/dx}\over{
x \ \cos \psi \ (d \psi/dx) }} 
\label{vpnearaxis}
\end{equation}
By using Eq.(\ref{parabpsiblofx}) for $\psi(x)$, 
Eqs.(\ref{Bernfarnearaxis}) and (\ref{vpnearaxis}), a simple differential equation for $a(x)$ is obtained, which is valid for
$( \pi/2 - \psi) \ll 1$. Its solution is given by Eq.(\ref{parabablofx}).
This provides the following  parametric representation of the solution 
very near the polar axis:
\begin{equation}
cos^2 \psi = {{\gamma Q_0 \rho_0^{\gamma - 2}(b) \ \mu_0 \alpha_0^2
}\over{
(\gamma - 1) \  \Omega_0^2 b^2}}
\left({{1}\over{x}} - {{1}\over{x^{2 - \gamma}  }} \right)
\label{parabpsiblofx} \end{equation}
\begin{equation}
 a = {{\gamma \ Q_0 \rho_0^{\gamma - 1} (b) \  \mu_0 \alpha_0
}\over{\sqrt{2} (\gamma - 1) \  \Omega_0^2 }}
\sqrt{ E_0 - {{\gamma}\over{\gamma - 1}} Q_0 \rho_0^{\gamma - 1}(b)}  \
\left( \ln{{{1}\over{x}}} - {{2-\gamma}\over{\gamma - 1}}\left(1 -
x^{\gamma - 1}\right)\right)
\label{parabablofx}\end{equation}

\subsection{Matching the Polar Boundary 
Layer Solution to the Outer Solution.}
\label{subsecmatchWKBpole}

Eliminating $x$,  when small (which corresponds to the 
outer regions of the polar boundary layer) between equations
(\ref{parabpsiblofx}) and (\ref{parabablofx}) 
gives the relation between $a$ and $\psi$ valid in these intermediate 
regions: 
\begin{equation}
\cos^2\psi = {{\gamma}\over{\gamma-1}}\ 
{{Q_0 \mu_0 \alpha_0^2 \rho_0^{\gamma -2}(b) }\over{ \Omega_0^2 \ b^2 }}
\ \exp \left( a \ \ 
{{( \gamma - 1) \ \sqrt{2} \ \Omega_0^2}\over{ \gamma \ 
Q_0 \rho_0^{\gamma -1}(b) \mu_0 \alpha_0 
\sqrt{   E_0- {{\gamma}\over{\gamma-1}} Q_0 \rho_0^{\gamma-1}(b)  }
}}
\right)
\label{outerlimofpsiofa} \end{equation}
It can be asymptotically matched to the outer solution
(\ref{solpsiofa}). In the vicinity of the polar axis, $\tan \psi$ is
large, and the constant term $\tan \psi(a_1)$ can be neglected if the
reference magnetic surface $a_1$ is not one of cylindrical geometry.
The relation (\ref{solpsiofa}) can be expressed as:
\begin{equation}
\cos \psi = \left( \cosh \chi \right) ^{-1}
\label{psicosh} \end{equation}
where 
\begin{equation}
\chi (a,b) =  {{1}\over{\sqrt{2} \mu_0 I_{\infty}(b)}}  \int_a^{a_1}
{{
\Omega(a')da'
}\over{
\sqrt{E(a') - I_{\infty}(b) \Omega(a') / \alpha(a')}
}}
\label{chiofa} \end{equation}
Near the polar axis
$\chi$ is large and approximately given by 
\begin{equation}
\chi(a,b) \approx {{1}\over{\sqrt{2} \mu_0 I_{\infty}(b)}} 
\left(  \int_0^{a_1}
{{\Omega(a')da'}\over{\sqrt{E(a') - I_{\infty}(b) \Omega(a') / \alpha(a')}}} 
- {{\Omega_0 \ a}\over{\sqrt{E_0 - I_{\infty}(b) \Omega_0  / \alpha_0 }}} 
\right)
\label{chismalla} \end{equation}
which, by Eq.(\ref{psicosh}), gives the inner 
limit of the outer solution as:
\begin{equation}
\cos^2\psi =
4 \ \ \exp \left( - \ \  {{\sqrt{2} }\over{ \mu_0 I_{\infty}(b) }}
\left[ \int_0^{a_1} 
{{ \Omega(a')da'}\over{ \sqrt{E(a') - I_{\infty}(b) \Omega(a') / \alpha(a')}}}
- {{ \Omega_0 \ a}\over{ \sqrt{E_0 - I_{\infty}(b) \Omega_0  / \alpha_0} }}
\right] \right)
\label{innerlimofpsiofaexpanded} \end{equation}

\subsection{Bennet Pinch Relation}
\label{Bennetaxis}

For the exponential arguments in Eqs.(\ref{outerlimofpsiofa}) and
(\ref{innerlimofpsiofaexpanded}) to coincide, it is necessary that:
\begin{equation}
{{\gamma}\over{\gamma -1}} \ Q_0 \rho_0^{\gamma -1} (b) = 
{{I_{\infty}(b) \  \Omega_0}\over{\alpha_0}} 
\label{bennetparab} \end{equation}
Eq.(\ref{bennetparab}) expresses a relation between the total current
supported by the polar boundary layer and its inner pressure.  The
existence of a relation between the total current and the central
pressure is common to all cylindrical plasma pinches and is usually
refered to as a Bennet relation.

\subsection{Polar Boundary Layer Current, Density and Radius}
\label{secondmatching}

For smooth asymptotic matching, the factors in front of the
exponential functions in Eqs.(\ref{outerlimofpsiofa}) and
(\ref{innerlimofpsiofaexpanded}) must coincide.  Taking
Eq.(\ref{bennetparab}) into account, this condition can be written as:
\begin{equation}
{{\gamma Q_0 \ \mu_0 \alpha_0^2 \rho_0^{\gamma - 2 }(b)
}\over{
(\gamma -1) b^2 \Omega_0^2 }}
=
4 \ \ \exp \left( - \
{{\sqrt{2} (\gamma -1) \ \Omega_0 }\over{\gamma \
\mu_0 \alpha_0 \ Q_0 \rho_0^{\gamma -1}(b) \
 }}  \
 \int_0^{a_1} {{ \Omega(a') \ da' }\over{ 
\sqrt{E(a') - I_{\infty}(b) \Omega(a') / \alpha(a') } 
}}
\right)
\label{rhozeroofb} \end{equation}
It can be more conveniently written by defining a length $\ell$, a
dimensionless measure of the density, $n_0(b)$, and a reference
magnetic flux $a_0$ by:
\begin{equation}
\ell^2 = {{\gamma}\over{(\gamma-1)}} \
{{Q_0 \rho_{A0}^{\gamma-1}}\over{\Omega_0^2}}
\qquad \qquad \qquad
n_0(b)  =  \rho_0(b) / \rho_{A0}
\qquad \qquad \qquad
a_0 =  {1 \over 2} \ \mu_0 \alpha_0 \sqrt{2 E_0} \ \ell^2
\label{normalparab} \end{equation}
and by using the notation
\begin{equation}
\lambda = \int_0^{a_1}
{{ \Omega(a') \  \sqrt{E_0} }\over{ 
\Omega_0 \ \sqrt{ E(a')- I_{\infty}(b) \Omega(a') / \alpha(a')} }} \ {{da'}
\over{a_0}}
\label{facteur} \end{equation}
The integral $\lambda$ depends on $\rho_0(b)$, or $n_0(b)$,
because of Eq.(\ref{bennetparab}).
The logarithm of Eq.(\ref{rhozeroofb}) then takes the form:
\begin{equation}
{{\lambda (n_0(b))}\over{n_0^{\gamma-1}(b) }} = 
(2 - \gamma) \ \ln (n_0(b))  \ +
\ \ln \left( {{4 b^2}\over{\ell^2}} \right)
\label{nzeroofRlog} \end{equation}
Since $n_0(b)$ is small and ($b/\ell$) large, an
approximate solution can be obtained by iteration,
initially neglecting the $\ln\left(n_0\right)$ term
on the r.h.s. as compared to $n_0^{-(\gamma -1)}$ on the l.h.s..
This gives at the simplest degree of
approximation:
\begin{equation}
n_0^{\gamma -1}(b) =  
{{\lambda(b) }\over{ 2 \ln \left( {{2 b}\over{\ell}}\right) }}
\label{nzeroofRsimple} \end{equation}
Eqs.(\ref{nzeroofRlog}) and (\ref{nzeroofRsimple}) are dominated by
the growth of the logarithm term on their r.h.s. They can be satisfied
for large $b$'s in two different ways, according to whether the
current $I_{\infty}(b)$ at the boundary layer's edge approaches a
finite value or decreases to zero.

When $I_{\infty}(b)$ approaches a
finite value, the Bennet relation (\ref{bennetparab}) shows that
the axial density should be independent of distance. The logarithmic
term in the denominator of equation (\ref{nzeroofRsimple}) should then
be compensated by a divergence of the numerator term $\lambda(b)$
(Eq.(\ref{facteur})). This shows that, as $b$ increases,
$I_{\infty}(b)$ should approach a limit that causes the integral on
the r.h.s. of Eq.(\ref{facteur}) to diverge. This implies that the
asymptotic limit of $I_{\infty}(b)$ be the absolute minimum of the
function $(\alpha E / \Omega)$, which, given the presence of a square
root denominator in the integral on the r.h.s. of Eq.(\ref{facteur}),
can only be approached from below.  Therefore $I_{\infty}(b)$ should
in this case asymptotically grow towards the absolute minimum of
$(\alpha E / \Omega)$. 

If $I_{\infty}(b)$ declines asymptotically to zero,
the function $\lambda(b)$ approaches a limit independent of $b$.
This indicates that equation (\ref{nzeroofRsimple})
should be satisfied by the decrease of $n_0(b)$ 
to zero at large distances. This is consistent with 
the fact that in this case all the magnetic surfaces 
flare out parabolically \citep{HeyvNorman89}
Equation (\ref{nzeroofRsimple}) implies a specific law of decrease of 
$\rho_0(b)$, namely,
denoting the limit of $\lambda(b)$
for zero current by $\lambda_0$: 
\begin{equation}
\rho_0(b) \approx \mu_0 \alpha_0^2 
\left(
{{\lambda_0 }\over{ 2 \ln \left( {{2 b}\over{\ell}} \right) }}
\right)^{{{1}\over{\gamma - 1}} }
\label{declineofrho0}
\end{equation}
It is then seen that $\rho_0(b)$ scales with $b$ as
$\left({\ln({{b}/{\ell}}})\right)^{-{{1}\over{\gamma-1}}}$.
The residual current $I_{\infty}(b)$, given by equation (\ref{bennetparab}),
slowly decreases as $\left({\ln({{2 b}/{\ell}}})\right)^{-1}$.

The radius $r_0$ of the circumpolar current channel at distance $b$
is the value of $r = b \ \cos \psi$ given by Eq.(\ref{parabpsiblofx})
for some intermediate value, of order unity, 
of the parameter $x$. This gives
\begin{equation}
r_0^2 = {{\gamma  Q_0 \rho_0^{\gamma - 2}(b) \ \mu_0 \alpha_0^2
}\over{
(\gamma - 1) \  \Omega_0^2}}
\label{radiusofpolarBL}
\end{equation}
For Poynting jets, this expression gives the specific value of 
this radius, since $\rho_0$ is given by Eq.(\ref{bennetparab}).  
For kinetic winds, the boundary layer radius
slowly increases with distance $b$ as 
$\left(\ln(b/\ell)\right)^{ {{(2 - \gamma)
 }\over{ 2(\gamma - 1)}} }$.

At this point the solution near the pole and in the 
field region extending in the polar-most cell is  completely 
determined. In particular, the solution for the flux distribution
in the field-region of this cell can, 
by using Eqs.(\ref{innerlimofpsiofaexpanded}),
(\ref{rhozeroofb}) and (\ref{chiofa}), be expressed as 
\begin{equation}
b^2 \cos^2\psi =
{{\gamma Q_0 \ \mu_0 \alpha_0^2 \rho_0^{\gamma - 2 }(b)
}\over{
(\gamma -1) \Omega_0^2 }}
\ \ \exp \left( 
{{\sqrt{2}}\over{ \mu_0 I_{\infty}(b)}}  \int_0^a
{{
\Omega(a')da'
}\over{
\sqrt{E(a') - I_{\infty}(b) \Omega(a') / \alpha(a')}
}}
\right)
\label{solinfieldaboutpole}
\end{equation}
and $\rho_0$ and $I_{\infty}$ are related by Eq.(\ref{bennetparab}). 

\subsection{Force-Free Polar Boundary Layers}
\label{appforcefreepolarbl}

If the poloidal magnetic pressure dominates plasma pressure, the
corresponding cylindrical structure is described by the transfield
mechanical balance equation (\ref{TFcurvass}):
\begin{eqnarray}
{r^2 \Omega \over \mu_0 \alpha I} {\vec{\nabla}a \over \vert \nabla a \vert
}\cdot
\vec{\nabla} \left({{\nabla a}^2 \over 2  \mu_0 r^2} \right)
+ {\Omega \over \alpha} {\vec{\nabla}a \over \vert \nabla a \vert }\cdot \vec
{\nabla} I = 0
\label{1Dforcefree} \end{eqnarray}
This equation should be associated with the cylindrical asymptotic 
form of Eq.(\ref{Bern})
\begin{equation}
E(a) = {{v_P^2}\over{2}} - {{r \Omega B_{\theta} }\over{\mu_0 \alpha}}
, \label{Bernassappend}
\end{equation}
with the mass conservation equation 
(\ref{defalpha}), with Eq.(\ref{rBthetaass}) and with Eq.(\ref{Bpversusa}). The latter
reduces in this geometry to
\begin{equation}
B_P = {1 \over r} \ {{da}\over{dr}}
\label{Bpversusacylapp} \end{equation}
The variables $v_P$ and $\rho$ can be
eliminated by Eqs.(\ref{defalpha}) and (\ref{rBthetaass}) in favour of
the field component $B_P$ and the quantity $I$. Using
Eq.(\ref{Bernassappend}), $B_P$ can then also be expressed in terms
of $I$. We are left, for given $b$, 
with a pair of ordinary differential equations 
for $I(r)$ and $a(r)$ where the first integrals, which are
known functions of $a$, also appear. This system is
\begin{equation}
{{d}\over{dr}} \left({{\mu_0 I^2}\over{r^4 \Omega^2}}
\ \left(E - {{I \Omega}\over{\alpha}}\right) \right)
+ {{ \mu_0 }\over{ r^2}} \ {{d}\over{dr}} \left( {{I^2}\over{2}} \right) = 0
\label{eqI1Dapp} \end{equation}
\begin{equation}
{{1}\over{r}} \ {{da}\over{dr}} = {{ \sqrt{2} \mu_0 I}\over{r^2 \Omega}}
\sqrt{E - {{I \Omega}\over{\alpha}}  }
\label{fluxdistr1Dapp} \end{equation}
As in the case of pressure-supported polar boundary layers, we
assume that the first integrals do not vary much over the current
carrying boundary layer and treat them as being constants, $\alpha_0$,
$E_0$ and $\Omega_0$ say.  This is valid for ${\Omega I}/{\alpha E}
\ll 1$ (app.{\ref{appfluxinbl}}) and allows to transform Eq.(\ref{eqI1Dapp}) 
to the simpler form:
\begin{equation}
{{E_0}\over{\Omega_0^2}} \ {{d}\over{dr}} \left({{ I^2}\over{r^4}}  \right)
+ {{ 1 }\over{ r^2}} \ {{d}\over{dr}} \left( {{I^2}\over{2}} \right) = 0
\label{eqI1Dsimpleapp}  \end{equation}
This shows that the natural unit length $r_0$ in the
force free pinch, or in other words the core radius of this pinch, is:
\begin{equation}
r_0^2 = {{E_0}\over{\Omega_0^2}}
\label{rzeroffpinchapp} \end{equation}
This  relation replaces equation (\ref{radiusofpolarBL}) 
for pressure-supported
axial boundary layers.
The solution of equation (\ref{eqI1Dsimpleapp}) is
\begin{equation}
I(r,b) = I_{\infty}(b)  \ {{\Omega_0^2 r^2}\over{
\Omega_0^2 r^2 + 2 E_0}}
\label{solpinchffapp}  \end{equation}
The solution (\ref{solpinchffapp}) can be asymptotically
matched with the field-region solution (\ref{solpsiofa}).  Actually
the solution (\ref{solpinchffapp}) itself explicitly shows this
continuuous transition from $I=0$ to $I= I_{\infty}$. 
It establishes the relation, which replaces Eq.(\ref{bennetparab}),
between the total electric current carried by the polar
boundary layer and the poloidal magnetic pressure that supports
the associated pinching force. 
The poloidal field
and plasma density can be expressed from 
Eqs.(\ref{defalpha}) and (\ref{rBthetaass})
in terms of $I$ as:
\begin{equation}
{{B_{P0}^2 (b)}\over{\mu_0}} = {{ 2 \mu_0 E_0}\over{\Omega_0^2}} \
\lim_{r \rightarrow 0}
\left( {{I^2}\over{r^4}} \right)
\label{Bpolatcenterffapp} \end{equation}
and
\begin{equation}
\rho_0(b) = {{\mu_0 \alpha_0}\over{\Omega_0}} \ \lim_{r \rightarrow 0}
\left( {{I}\over{r^2}} \right)
\label{rhocentralffapp} \end{equation}
The limiting  value involved is obtained from Eq.(\ref{solpinchffapp}):
\begin{equation}
\lim_{r \rightarrow 0} \left( {{I}\over{r^2}} \right) =
{{I_{\infty} \Omega_0^2}\over{2 E_0}}
\label{jcentralffapp} \end{equation}
Equations (\ref{Bpolatcenterffapp}), (\ref{rhocentralffapp}), (\ref{jcentralffapp})
can be synthesized in the relation:
\begin{equation}
{{B_{P0}^2(b)}\over{\mu_0 \rho_0(b) }} = {{ I_{\infty}(b) \Omega_0}\over{ \alpha_0}}
\label{Bennetffapp}  \end{equation}
which is the form of Bennet pinch relation appropriate to this case.
It differs from Eq.(\ref{bennetparab}) 
only by the substitution of the axial Alfv\'en speed to the
axial sound speed.

\section{Null Surface Boundary Layers}
\label{nullsurfaceBL}

\subsection{Divergence of the 
Mass Flux to Magnetic Flux Ratio at Null Surfaces}
\label{subsecdivalpha}

At a null surface, labelled by $a_n$ say, the mass flux to
magnetic flux ratio, $\alpha (a)$, defined by
Eq.(\ref{defalpha}), diverges as $a \rightarrow a_n$ if
there is mass flux on this null surface.  We
show in appendix (\ref{appalphaatnullsurf}) that the divergence of
$\alpha(a)$ is as
\begin{equation}
\alpha(a) \sim  {{1}\over{\vert a_n -a \vert^{\nu} }}
\label{divalpha} \end{equation}
where $\nu$ is a positive number stricly less
than unity. Usually, $\nu = 1/2$. 

\subsection{Structure of Null Surface Boundary Layers.}
\label{appnonequatnullsurf}

The structure of the flow near a null-surface
can be derived from  the
transfield equation (\ref{TFcurvass}). 
The ratio of the gas pressure to the toroidal magnetic pressure
decreasing to zero as the axial distance $r$ increases,
the thickness of the pressure-dominated region
about the null surface, $\eta$, becomes small at large distances.
The gradient operator normal to the magnetic surface,
$\vec{\nabla}a \cdot \vec{\nabla}/\vert \nabla a \vert$,
noted as $- \vec{n} \cdot \vec{\nabla}$, 
is then of order $\eta^{-1}$
when acting on $I$ or on $Q \rho^{\gamma}$. 
The left hand side of equation (\ref{TFcurvass})
being smaller than ($v_P^2/r$) (see section (\ref{dominantforces})), can
thus be neglected. Similarly the variable $r$ can be treated as a constant
because ($\vec{n} \cdot \vec{\nabla} r$),
of order unity, is negligible to ($r/\eta$). Eq.(\ref{TFcurvass})
then reduces, using Eq.(\ref{IofBtheta}), to:
\begin{equation}
\vec{n} \cdot \vec{\nabla} \left( Q \rho^{\gamma} +
{{B_{\theta}^2}\over{2 \mu_0 }} \right) = 0
\label{presseqatnullapp} \end{equation}
which, by  using Eq.(\ref{rBthetaass})
and regarding $Q$ and $\Omega$ as 
constants, $Q_n$ and $\Omega_n$, takes the form
\begin{equation}
\vec{n} \cdot \vec{\nabla} \left( Q_n \rho^{\gamma} +
{ 1 \over 2} \ {{\rho^2 r^2 \Omega_n^2 }\over{\mu_0 \alpha^2}} \right) = 0
\label{eqrhoatnull} \end{equation}
Since $\alpha$ diverges 
at a null surface (section \ref{subsecdivalpha} and app.
\ref{appalphaatnullsurf}) this integrates as
\begin{equation}
Q_n \rho^{\gamma} + { 1 \over 2}
\ {{\rho^2 r^2 \Omega_n^2 }\over{\mu_0 \alpha^2}} =
Q_n \rho_n^{\gamma} (r)
\label{pressequilnull}  \end{equation}
where $\rho_n (r)$ is the density at the cylindrical distance  $r$ of
the axis on this null surface. Let us introduce the parameter
\begin{equation}
X = \rho /\rho_n(r)
\label{Xneutralgeneral}
\end{equation}
The mass to flux ratio $\alpha(a)$ is given in terms of $X$ 
by Eq.(\ref{pressequilnull}), namely:
\begin{equation}
{{1 }\over{\mu_0 \alpha^2(a) }}
= {{2 Q_n \rho_n^{\gamma - 2}(r)}\over{\Omega_n^2 r^2}}
\left({{1 }\over{X^2}} - {{1}\over{ X^{2 - \gamma} }} \right)
\label{alphanullparam} \end{equation}
Note that, as it should, $\alpha^{-1}$ vanishes as the neutral
surface is approached, when $X$ approaches unity. 
When the functional dependence of  $\alpha(a)$ on $a$
is known, Eq.(\ref{alphanullparam}) gives
$a$ in terms of $X$. Since this dependence is however not
universal (see Eq.(\ref{divalpha})), this step cannot
be performed in a general way. Close to a neutral surface,
the Bernoulli equation  (\ref{Bern}) reduces, 
in the asymptotic domain, to :
\begin{equation}
E = {{v_P^2}\over{2}} + {{\gamma }\over{\gamma -1}} 
Q \rho^{\gamma -1} - {{ r \Omega B_{\theta} }\over{ \mu_0 \alpha}}
\label{bernnearneutralfar}
\end{equation}
where $B_{\theta}$ can be obtained from Eq.(\ref{rBthetaass}).
Using Eq.(\ref{alphanullparam}) to account for the variation
of $\alpha (a)$  in the neighborhood of the neutral surface,
we obtain, using Eq.(\ref{Xneutralgeneral}) and neglecting the
variations of $r$ accross the boundary layer, 
\begin{equation}
{{v_P^2}\over{2}} \approx E_n - {{2 - \gamma}\over{\gamma -1}} Q_n 
\rho_n^{\gamma -1} X^{\gamma -1} - 2 Q_n \rho_n^{\gamma -1} X^{-1}
\label{vpsquarenearneutralfar}
\end{equation}
On the other hand, $\vert \vec{v}_P \vert$ is related 
to $\vert \vec{\nabla} a \vert$ by Eqs.(\ref{Bpversusa}) and (\ref{defalpha}).
Using Eq.(\ref{gradaversuspsi}), this provides
a differential equation for the dependance of the latitude angle 
$\psi$ on $a$, or equivalently, on $X$, at a given $b$.
Using Eq.(\ref{vpsquarenearneutralfar}) 
this differential equation can be written as:
\begin{equation}
b \ d\psi = - {{\alpha(a) da }\over{ \sqrt{2} \ r X \rho_n
 \ \sqrt{ E - {{2 - \gamma}\over{\gamma -1}} Q_n \rho_n^{\gamma -1} X^{\gamma
-1} -
2 Q_n \rho_n^{\gamma -1} X^{-1}    }    }}
\label{psinullBLofX} \end{equation}
Eq.(\ref{psinullBLofX}) could be brought to quadratures for $\psi(X)$
if the explicit dependance of $a$ on $X$ could be deduced from
Eq.(\ref{alphanullparam}) for known, and invertible, $\alpha (a)$,
such as for example in Eq.(\ref{divalpha}).  This step will not be
taken here explicitly because the relation between $a$ and $\alpha$
does not have an universal character, even near a neutral
layer. Although we have not derived the solution explicitly, we can
still proceed to deduce the necessary conditions for a smooth matching
to the solution in the far field.

\subsection{Matching the Null Surface
Boundary Layer Solution to the Field}
\label{matchequatortoexternal}

The solution in the boundary layer about the neutral surface
is now given by
equations (\ref{alphanullparam}) and (\ref{psinullBLofX}). It
can be asymptotically matched to the field-region solution
which is expressed in
differential form by equation (\ref{eqpsiofageneral}).
In the field-region near the
neutral surface Eq.(\ref{eqpsiofageneral}) reduces to:
\begin{equation}
d \psi = - {{\Omega_n da}\over{ \mu_0 I_{\infty}(b) \sqrt{2}
\sqrt{E_n - I_{\infty}(b) \Omega_n / \alpha(a) } }}
\label{psiinfieldnearequa}  \end{equation}
On the other hand, eliminating $X$ between
Eqs.(\ref{alphanullparam}) and (\ref{psinullBLofX})
in the small $X$ limit, which is relevant
to the outskirts of the equatorial boundary layer,
we obtain a differential equation for $\psi$ valid  in this region:
\begin{equation}
d \psi = - {{\Omega_n da}\over{ r \sqrt{2}
\sqrt{2 \mu_0 Q_n \rho_n^{\gamma}(r) }
\sqrt{ E_n - {{\Omega_n}\over{\mu_0 \alpha}}
\sqrt{2 \mu_0 Q_n r^2 \rho_n^{\gamma}(r)
} } }}
\label{psiinouterequaBL}  \end{equation}
Matching requires that equations (\ref{psiinfieldnearequa}) and
(\ref{psiinouterequaBL}) be identical. This implies that the total
current at the edge of the null surface boundary layer at a distance
$b$ from the wind source (corresponding to an axial distance $r$) is
related to the density at the center of the layer by:
\begin{equation}
\mu_0 I_{\infty} (r) = \sqrt{ 2 \mu_0 Q_n r^2 \rho_n^{\gamma}(r) }
\label{bennetnull} \end{equation}
This relation expresses the
balance between gas pressure at the null surface
and the magnetic pressure just at the outer edge of its boundary layer,
as
expected for a sheet-pinch.
As a result we find that, for Poynting jets, the equatorial density decreases as
\begin{equation}
\rho_n(r) \sim r^{- {2\over \gamma}}
\label{assrhonullcylind} \end{equation}
For kinetic winds,
$I_{\infty}(r)$ decreases as $1/ (\ln \left(2 r/\ell \cos \psi_n)\right))$ and
the
density at the null magnetic surfaces declines as:
\begin{equation}
\rho_n(r) \sim
\left( r \  \ln \left({{ 2 r
}\over{
\ell  \cos \psi_n}}\right) \right)^{- {2\over \gamma}}
\label{assrhonullparab}  \end{equation}

\subsection{Flux and Current Distribution Near a Neutral Surface.}
\label{subsecfluxversuslatitudeatneutral}

Eq.(\ref{eqpsiofageneral}) indicates that $(d\psi/da)$ becomes
infinite at a neutral magnetic surface, where $I$ vanishes.  Does
that imply that $\tan \psi$ becomes infinite at neutral surfaces?  In
a given current cell, Eq.(\ref{eqpsiofageneral}) integrates similarly
to Eq.(\ref{solpsiofa}) to:
\begin{equation}
\tan (\psi (a,b)) = \tan(\psi (a_1, b))
+ \sinh \left( \int_a^{a_1}
{{1}\over{\sqrt{2} \mu_0 I(a',b) }} \
{{\Omega(a') da'}\over{
\sqrt{ E(a') - I(a',b) \Omega(a') / \alpha(a') }
}} \right)
\label{solpsiofanearneutral}
\end{equation}
Whether $tan (\psi)$ diverges or not as $a$ approaches $a_n$ depends on 
the behaviour of the integral on the r.h.s. of Eq.(\ref{solpsiofanearneutral}) as
the neutral surface is approached and this in turn depends on how $I(a,b)$
varies with $(a - a_n)$ as the neutral surface $a_n$ is
approached. This can be deduced from the solution expressed by
Eq.(\ref{alphanullparam}).  The current $I$ in the
asymptotic domain is given by Eq.(\ref{Iasympt}). Using
Eqs.(\ref{Xneutralgeneral}) and (\ref{alphanullparam}) it is found
that:
\begin{equation}
I^2 \approx {{2 Q_n  r^2  \rho_n^{\gamma}(r)}\over{\mu_0}}
\left( 1 - X^{\gamma} \right)
\label{Inearnull}
\end{equation}
The parameter $X$ is related to $\alpha$, or $(a - a_n)$,
by Eq.(\ref{alphanullparam}). It is shown in
section (\ref{subsecdivalpha}) that $\alpha$
scales with $(a_n - a)$ as:
\begin{equation}
\alpha^{-1} \sim (a_n -a)^{\nu}
\label{alphascalingatneutral}
\end{equation}
where $\nu$ is a positive exponent strictly smaller than unity, usually
equal to ${1 \over 2}$.  Very near the neutral surface, 
$X$ is close to unity. From Eq.(\ref{alphanullparam}),
it is found that:
\begin{equation}
 (a -a_n)^{2\nu} \sim {{2 Q_n \rho_n^{\gamma -2}}\over{\Omega_n r^2 X^2}}
\left( 1 - X^{\gamma} \right)
\label{ascalingwithX}
\end{equation}
Comparing this with Eq.(\ref{Inearnull}), we conclude that,
very near the null surface,
\begin{equation}
I \sim ( 1 - X^{\gamma})^{1/2} \sim (a -a_n)^{\nu}
\label{Iscalingwithaatneutral}
\end{equation}
With such a dependence of $I$ on $a$ the integral in
Eq.(\ref{solpsiofanearneutral}) converges as $a$ approaches $a_n$,
since $\nu$ is less than unity. Therefore, neutral surfaces do not
become vertical when approached from a conical region.

\section{Shape of the Magnetic Surfaces}
\label{sectionshape}

We have now obtained a complete solution 
in the asymptotic domain, both in field-regions
(section (\ref{subsecfluxdistrgeneral})) near 
the pole (section (\ref{solutionpolarBL})) and near neutral 
magnetic surfaces (section (\ref{appnonequatnullsurf})).
Since the integral in Eq.(\ref{solpsiofanearneutral}) 
converges at a neutral surface, integration may be
started at the equator, irrespective of whether or not it is a neutral
surface. At the equator, $\tan \psi$ vanishes,
and thus the integration constant 
of Eqs.(\ref{solpsiofa}) and (\ref{solpsiofanearneutral}) 
can be taken to be zero. The solution is then extended to other angles $\psi$ 
by using either Eq.(\ref{solpsiofanearneutral}) or 
appropriate solutions in the neutral or polar boundary layers
to specify how $I(a,b)$ depends on $a$ at given $b$.

The dependence of $I_{\infty}(b)$ on $b$ in field-regions
is determined by solving  Eq.(\ref{nzeroofRlog}), 
$\lambda(b)$ being defined by Eq.(\ref{facteur}) and $I_{\infty}$ by 
Eq.(\ref{bennetparab}).  Eq.(\ref{nzeroofRlog})
may have one or two solutions at large $b$, 
depending on the 
function $(\alpha E /\Omega)$. 

We now have gathered all the information needed
to calculate the asymptotic shape of magnetic surfaces,
both in the field-region and in the various boundary layers.

\subsection{Magnetic Surfaces in Field-regions of Poynting Jets}
\label{subsecshapefieldPoyntingjets}

When the flow is a Poynting jet, 
its magnetic structure  consists of cylindrical 
surfaces nested into conical ones.
The relation between flux and radius  for cylindrical surfaces
is given by equations (\ref{parabpsiblofx}) and (\ref{parabablofx})
in the polar boundary layer, noting that
$r \equiv b \cos \psi$, and  by (\ref{solinfieldaboutpole})
and  (\ref{bennetparab}) outside of it. 
The dependence of  the latitude angle of conical magnetic surfaces
on flux is given by (\ref{solpsiofa}) in the field-region.
When there is only one cell extending from pole to
equator, $a_1$ can be taken as the equatorial flux $A$ 
and $\tan \psi (a_1)$ reduces to zero, since 
there is vanishing flux left in the equatorial 
boundary layer at large distances. The shape of the magnetic surfaces 
in the equatorial boundary layer itself 
is obtained in section \ref{subsecshapeequator}.

\subsection{Magnetic Surfaces in Field-regions of Kinetic Winds}
\label{subsecshapefieldkinwinds}

The magnetic surfaces of kinetic winds are described
by Eq.(\ref{solpsiofa}), $I_{\infty}(b)$ being now given 
by the analysis of section (\ref{secondmatching})
(Eqs.(\ref{nzeroofRlog})-(\ref{nzeroofRsimple})). By 
Eq.(\ref{bennetparab}) we get, for small $I_{\infty}(b)$:
\begin{equation}
\mu_0 I_{\infty}(b) = 
\left( \int_0^A {{\Omega (a) da}\over{\sqrt{2E(a)} }} \right)
\ \ {{1}\over{\ln \left({{b}\over{l}}\right)}}
\label{Iinftyexplicitlog} \end{equation}
We consider the case of polar and equatorial boundary layers of kinetic winds
in sections \ref{subsecshapepolbl} 
and \ref{subsecshapeequator}
respectively. 
In the field-regions the
shape of magnetic surfaces is given  by the solution 
of the following differential equations
for $r(b)$ and $z(b)$:
\begin{equation}
dr = \cos\psi(a, b) \ \ db  \qquad \qquad dz = \sin\psi(a, b) \ \ db
\label{fieldlineeq} \end{equation}
which are to be integrated in $b$ for constant $a$. 
The latter argument will be omitted below.
The angle $\psi(a,b)$ is given by 
Eqs.(\ref{psicosh})-(\ref{chiofa}). 
Since magnetic surfaces are in this case parabolic, $\psi$ 
is close to $\pi/2$ and the spherical distance $b$
can be identified with $z$. Eq.(\ref{fieldlineeq}) then reduces to 
\begin{equation}
dr = \cos \psi(z)  \ dz
\label{fieldlineeqparaxial} \end{equation}
Similarly, Eq.(\ref{psicosh}) for $\psi$ simplifies, for large $\chi$'s 
and for $I_{\infty}(b)$ as given by eq. (\ref{Iinftyexplicitlog}), to: 
\begin{equation}
\cos\psi(z)  = \ 2 \left({{\ell}\over{2 z}}\right)^{k(a)}
\label{cospsipowerlaw} \end{equation}
where $k(a)$ is:
\begin{equation}
k(a) = {{    \int_a^A
	{{\Omega(a') \ da'}\over{\sqrt{E(a')}}} }\over{
\int_0^A
	{{\Omega(a') da' }\over{\sqrt{E(a')} }}  }}
\label{cospsiexponent} \end{equation}
Noting $q(a) = 1 - k(a)$, the solution of Eq.(\ref{fieldlineeqparaxial}) is:
\begin{equation}
{{r}\over{\ell}} = {{1}\over{q(a)}}
\left({{2 z}\over{\ell}}\right)^{q(a)}
\label{paraboloidshape} \end{equation}
The magnetic surfaces are then, in the paraxial region 
outside of the polar boundary
layer, a collection of nested power-law paraboloids of variable exponent. 
Some sections of the magnetic surfaces, though 
extending out of the equatorial boundary layer, may still be close enough
to the equator that the paraxial approximation $\psi \approx \pi/2$,
is inappropriate for them.
Their shape should be found by integrating
Eqs.(\ref{fieldlineeq})
with no further approximation,
as done in appendix \ref{appgeneralparaboloids}.

\subsection{Magnetic Surfaces in the Polar Boundary Layer}
\label{subsecshapepolbl}

In the polar boundary layer of Poynting jets, magnetic surfaces are cylinders.
For kinetic winds, 
the paraxial approximation is fully justified in this region and
the shape of magnetic surfaces is given by 
Eqs.(\ref{parabpsiblofx})-(\ref{parabablofx}), 
with the notations of Eqs.(\ref{normalparab}). 
Eq.(\ref{parabpsiblofx}) can be written as:
\begin{equation}
{{r^2}\over{\ell^2}} = {{1}\over{n_0^{2-\gamma}(b)}}
\left({{1}\over{x}}-{{1}\over{x^{2-\gamma}}}\right)
\label{psiofxagain} \end{equation}
whith Eq.(\ref{parabablofx}) 
providing the value of the parameter $x$ in terms of $a$ by:
\begin{equation}
\ln{{1}\over{x}} - {{2-\gamma}\over{\gamma-1}}
\left(
{1-x^{\gamma-1}}
\right) =
{{ a }\over{ a_0 }} \  {{1}\over{ n_0^{\gamma-1}(b)}}
\label{xrelatedtoainbl} \end{equation}
The dimensionless axial density $n_0(b)$ is 
approximately given  by Eq.(\ref{nzeroofRsimple}). 
The spherical distance to the origin being almost identical to $z$,
Eqs.(\ref{psiofxagain})-(\ref{xrelatedtoainbl}) constitute  
a set of coupled equations relating $x$, $r$ and $z$.
With the notations of Eqs. (\ref{normalparab})-(\ref{facteur})  they can
be written as:
\begin{equation}
{{r^2}\over{ \ell^2}} =
\left( {{1}\over{x}} - {{1}\over{x^{2-\gamma}}} \right)
\left( 2 {{   \ln \ ({{2 z}/{\ell}}) }\over{   \lambda  }}
\right)^{{2-\gamma}\over{\gamma-1}}
\label{rofxandzinbl} \end{equation}
\begin{equation}
\left( {{ 2a}\over{ \lambda  a_0}} \right)  \ \ln  \left({{2 z}\over{\ell}}\right) = 
\left(\ln {{1}\over{x}} -{{2-\gamma}\over{\gamma-1}}
\left(1-x^{\gamma-1}\right)\right)
\label{zofxandainbl} \end{equation}
This system  can be solved to give $r$ and $z$ in terms of $x$.
For small $x$ we recover 
the paraxial field-region solution, while for  $x$ close
to unity we obtain the shape of magnetic surfaces in 
the pressure-dominated region very near the
polar axis. In this region the magnetic  surfaces 
switch from algebraic paraboloids to exponential ones, 
their shape being given by:
\begin{equation}
{{r}\over{\ell}} = \sqrt{ {{a}\over{ a_0}}  } \
\left( {2 \over \lambda} \right)^{{1}\over{2(\gamma-1)}}\
\left({\ln \ {{2 z}\over{\ell}}}\right)^{{1}\over{2(\gamma-1)}}
\label{shapeincenterofbl} \end{equation}

\subsection{Magnetic Surfaces in the Equatorial Boundary Layer}
\label{subsecshapeequator}

Let us assume for simplicity that the only neutral
surface is the equatorial plane.
The information on the shape of magnetic surfaces in its boundary layer
is provided by the parametric solution of
Eqs.(\ref{alphanullparam}) and (\ref{psinullBLofX}). The density 
in the equatorial plane at the distance $r \equiv b$
from the source, $\rho_e(b)$,
is related to the polar residual current $I_{\infty}(b)$ 
by Eq.(\ref{bennetnull}). $I_{\infty}(b)$  approaches a constant 
value for Poynting jets  
and decreases logarithmically for kinetic winds. 
Both cases can be unified by writing 
\begin{equation}
I_{\infty}(b) = {{J_m}\over{ \left(\ln (2 b/\ell) \right)^m }}
\label{behaveIinfty} \end{equation}
where $m =0$ for Poynting jets and $m=1$ for kinetic winds, 
the factor $J_m$ being different in the two cases. 
Equation (\ref{bennetnull}) then gives for the equatorial density
\begin{equation}
\rho_e(b) = \left({{\mu_0 J_m^2}\over{ 2 Q_e}} \right)^{1/\gamma}
\ \left( {{1}\over{ b \ (\ln (2 b/\ell))^m }}  \right)^{2/\gamma}
\label{behaverhoequat} \end{equation}
For small $X$ we recover from Eqs. (\ref{alphanullparam}) and
(\ref{psinullBLofX}) the results valid in the
field-region. Eliminating $\alpha$ from Eq.(\ref{psinullBLofX}) and
using Eqs.(\ref{alphanullparam}) and (\ref{bennetnull}) we get:
\begin{equation}
z =  b \ 
\int_a^A {{ \Omega_e \ da}\over{
\sqrt{2 \mu_0 Q_e b^2 \rho_e^{\gamma}(b)} \ \ \sqrt{2 \  ( E_e - I_{\infty}(b) \Omega_e /\alpha)} 
}}
\label{zofRequatoroutskirt} \end{equation}
When $I_{\infty}$ approaches a non-vanishing constant
so does, from Eq.(\ref{bennetnull}), $b^2 \rho_e^{\gamma}(b)$. 
Then $z/b$ approaches, at fixed $a$,  a  constant value: 
the magnetic surfaces become conical at the outskirts of the
equatorial boundary layer, as they should.
For kinetic winds, Eq.(\ref{zofRequatoroutskirt}) gives, 
considering Eq.(\ref{behaveIinfty}):
\begin{equation}
z \approx   {{\Omega_e (A -a)}\over{ J_0 \sqrt{2 E_e} }} \  \
b \ln \left( {{2 b}\over{\ell}} \right) 
\label{shapeoutskirtequaparab} \end{equation}
and the magnetic surfaces become slightly convex paraboloids
at the outskirts of the
equatorial boundary layer.
By contrast, in the region of the equatorial boundary layer 
where the gas pressure dominates,  $X$ is close to unity and
the integration of Eqs.(\ref{alphanullparam}) and (\ref{psinullBLofX})
gives: 
\begin{equation}
z = \int_a^A {{\alpha(a) da}\over{ b \ \rho_e(b) }}
{{1}\over{ 
\sqrt{2  \left( E_e - {{\gamma }\over{\gamma -1}} \ Q_e \rho_e^{\gamma -1}(b) \right)} }} 
\label{zofRequatorclose} \end{equation}
where again $\rho_e(b)$ is given by Eq.(\ref{behaverhoequat}).
It is then found that $z(b)$ scales, at fixed $a$, as
\begin{equation}
z \sim b^{ {{2 - \gamma}\over{\gamma}} }
\label{zofRinequatBLcyl} \end{equation}
for Poynting jets and as
\begin{equation}
z \sim b^{ {{2 - \gamma}\over{\gamma}} } \ 
\left(\ln \left( {{2 b}\over{\ell}} \right)  \right)^{2 \over \gamma}
\label{zofRinequatBLparab}  \end{equation}
for kinetic winds.
Note that in this region magnetic surfaces 
are concave and bend towards the equator.
This agrees with the conclusions of \citet{okamoto99},
although the force balance considered by
this author is between 
the Lorentz force and the poloidal curvature inertia force, whereas 
we consider balance between the Lorentz force
and the pressure gradient force.
This does not contradict the results of \citet{HeyvNorman89} because
this is by no means the terminal shape of these magnetic surfaces.
Indeed, as discussed in section \ref{subsecexit} below and in 
Appendix \ref{appangleofexitfromeqbl}, any magnetic field line eventually escapes
the equatorial boundary layer region, first joining a region at
its outskirts where its shape becomes
conical or parabolic as indicated by Eq.(\ref{zofRequatoroutskirt}) 
and eventually reaching the field-region. 

\subsection{Exit from the Equatorial Boundary Layer}
\label{subsecexit}

When magnetic surfaces exit the equatorial boundary layer, they do so
with an angle to the equatorial plane that decreases with distance.
The discussion of section (\ref{subsecraytracing}) assumed trajectories
orthogonal to magnetic surfaces to cross normal to the equator. For
dipolar symmetry, an equatorial boundary layer is always present
between the field-region and the equator itself. The boundary
conditions used in section (\ref{subsecraytracing}) are thus consistent
only if the latitude angle of field lines at the outer edge of the
equatorial boundary layer becomes increasingly negligible with
distance. We show in appendix \ref{appangleofexitfromeqbl}
that the slope $(\partial z/\partial b)$ of the exiting field line
indeed decreases 
to zero with distance along the equatorial
boundary layer. 

\subsection{Justification of WKBJ Treatment}

Our WKBJ treatment of the field-region solution is valid only if
the inertia force associated with the curvature of the poloidal
motion remains negligible. In the case of Poynting jets, the poloidal
field lines in the field-region asymptotically become exactly straight
so that solving Eq.(\ref{TFcurvass}), supposedly valid to order
$r_A/r$, while ignoring the curvature term at the l.h.s. of
Eq.(\ref{TFcurvform}) is obviously consistent. In the case of kinetic
winds, the poloidal field lines are described
by Eqs.(\ref{zofRgeneral})-(\ref{rofRgeneral}) of
appendix \ref{appgeneralparaboloids}, where their radius of
curvature has also been calculated. This radius is
proportional to $(r^{{{1+k}\over{1 -k}}}/k)$ 
on field lines for which the exponent defined 
by Eq.(\ref{cospsiexponent}) is $k$.
Since $k$ approaches unity at the pole, the neglect of the force due to
poloidal curvature is fully justified in these regions. Near the
equator, where $k$ approches zero, the curvature radius comes closer
to scaling as $r$, but remains still much larger than $r$. 
This is because $k$ never reduces exactly to zero and because 
of the presence of the dividing factor $k$, which reflects
the fact that when field lines become tangent to the equator, their
curvature must be very small.

\section{Conclusions}
\label{secconclusion}

We have derived global solutions for the asymptotic structure of
non-relativistic, rotating, stationary, axisymmetric, polytropic, unconfined,
perfect MHD winds.  The five lagrangian first integrals are assumed to
be known.

The asymptotic structures have been found to consist of vast regions,
called field regions, which are devoid of any significant residual
poloidal electric current density. Residual current flows in thin regions 
in the vicinity of the polar axis and the
neighbourhood of null magnetic surfaces. Null surfaces can occur at
polarity reversals of the wind source or extend 
over dead zones. They delineate separate cells in which the poloidal 
current achieves closure.

For kinetic-energy-dominated winds the conversion of total wind energy
to kinetic energy is shown to progress only logarithmically with
distance.

All winds have been shown to possess a circumpolar current-carrying
boundary layer, which has the structure of a pressure-supported
plasma-jet pinch. Null-surface boundary layers have the
structure of pressure-supported current sheets. The total electric
current is constant or slowly diminishes with distance according to an
inverse logarithmic law for the Poynting flux and kinetic winds
respectively. This dimunition is caused by minute amounts of current
flowing through the diffuse field regions from the pole to the nearest
null surface.

The pressure in the center of these regions, where the toroidal field
vanishes, is related to the residual current by Bennet pinch
relations. The plasma density remains constant at the polar axis or
declines as a negative power of the logarithm of the distance to the
wind source as above. Therefore, Poynting flux can be retained, even in
kinetic winds, over large distances.

We have calculated the structure of the flow in all possible regions
including field regions, the polar boundary layer and null-surface
boundary layers. The solution is given in terms of standard
first-integrals using a WKBJ approximation that incorporates the weak
dependence on the distance from the source. A complete solution has
been constructed by asymptotic matching of these separate pieces of
the solution. Global relations are found between the circumpolar  current
and the density at the polar axis or at neutral surfaces 
(Eqs.(\ref{bennetparab}) and (\ref{bennetnull})). 
We have established similar 
relations in the case of jets with force free polar 
boundary layers (Eq.(\ref{Bennetffapp})).

\begin{figure}
\epsscale{1}\plotone{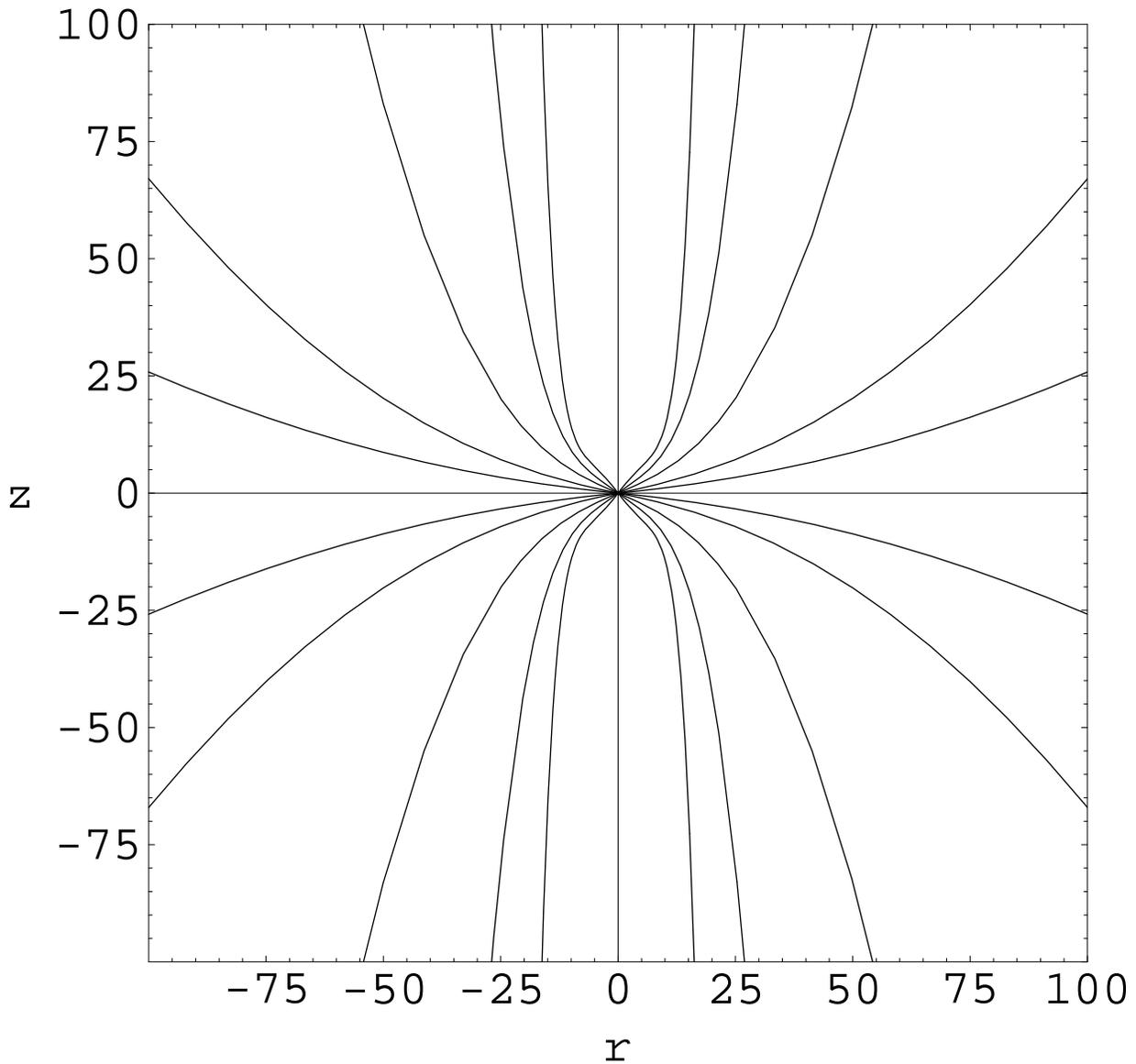}
\caption{The magnetic field structure for kinetic winds
in the asymptotic field region, from Eq.(\ref{paraboloidshape}).
The functions $E(a)$ and $\Omega(a)$ have been
taken as constant which implies that $q(a) = (a/A)$. On
the pole $q(0)$ = 0 and at the equator $q(a)$ = 1.
The field lines in each quadrant correspond to
$a/A = $ $0.2$, $0.4$, $0.6$, $0.8$, $0.9$ resp. An interpolation formula
has been used to connect the asymptotic solution to a split-monopole
field at the origin. The scale for $r$ and $z$ is arbitrary.
\label{fig3}}
\end{figure}

The shapes of magnetic surfaces in all parts of the solution and in
all relevant regimes have been calculated as well. The results are as
follows:

\noindent
I. For winds which are kinetic-energy dominated at infinity:

\noindent
(i) In the free field, the magnetic surfaces focus into algebraic
paraboloids (Eq.(\ref{paraboloidshape})) as shown in Fig.\ref{fig3}.

\begin{figure}
\epsscale{1}\plotone{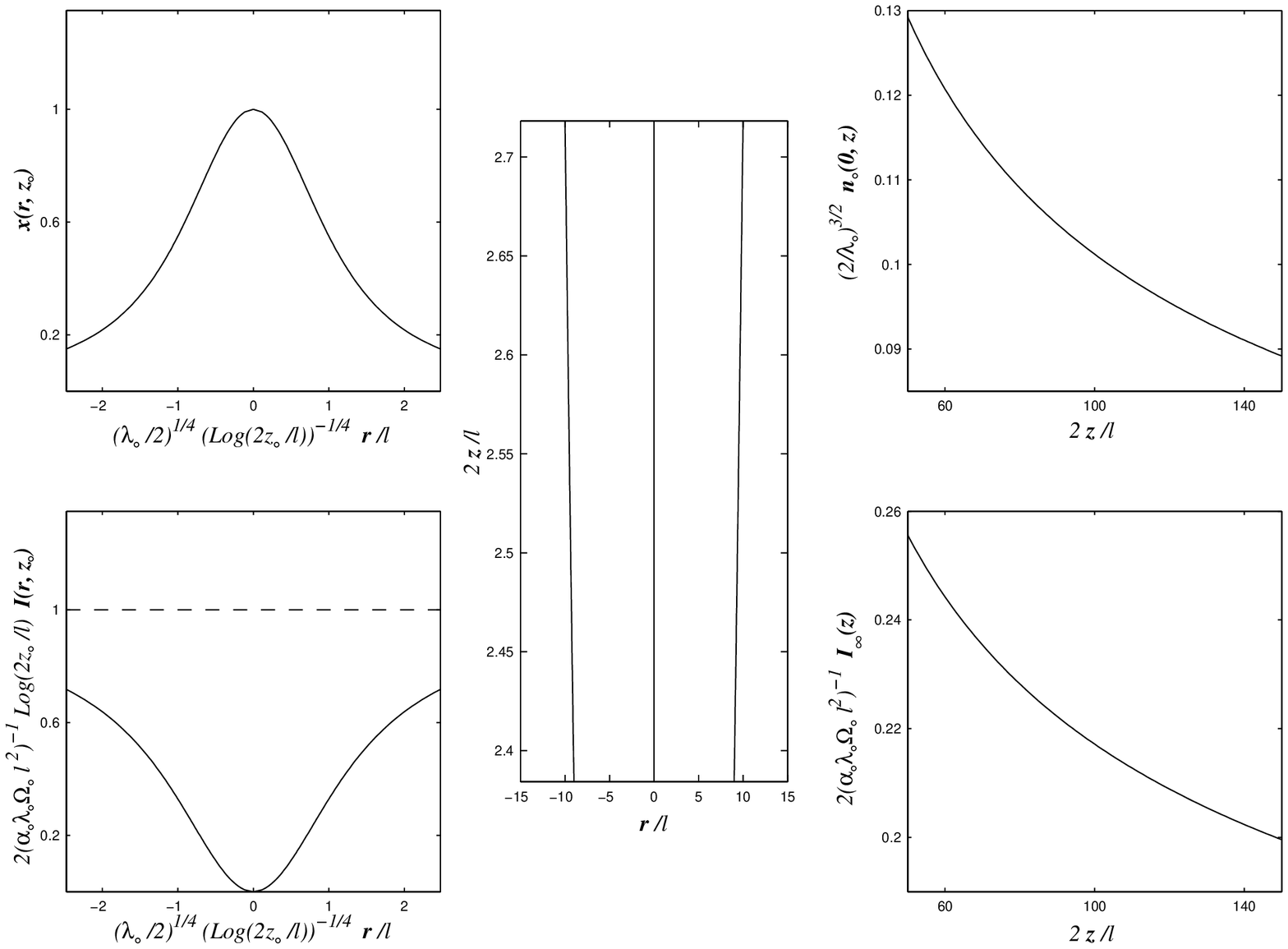}
\caption{
The behaviour of kinetic winds near the polar axis. The panels show:
magnetic field structure (central frame); the normalized 
density (Eq.\ref{normalparab}) at the polar
axis vs. the distance along the polar axis (upper right); 
the total current integrated about the polar 
axis vs. the distance along the polar axis (lower right); 
the ratio of the density to its polar value
at the same $z$ across the polar pinch (upper left) and the integrated current
across the polar axis (lower left). The latter two quantities are plotted for 
some arbitrarily chosen $z_0$. 
The central frame uses a value for $a$ such that the
proportionality constant in Eq.(\ref{shapeincenterofbl})
becomes unity. The slightly parabolic shape of the field lines is not clearly
visible on the scale of the plot. The right panel 
curves are from Eqs.(\ref{bennetparab}), (\ref{normalparab}) and (\ref{nzeroofRsimple}) with
$\lambda_0$ given by Eq.(\ref{facteur}) 
for negligibly small $I_\infty$. The left panel curves are 
from Eqs.(\ref{defdensratio}) (\ref{parabpsiblofx}), 
(\ref{parabablofx}), 
(\ref{Iasympt}), (\ref{normalparab}) and (\ref{declineofrho0}). The
adiabatic index here is $\gamma = 5/3$.
\label{fig4}
}
\end{figure}

\noindent
(ii) In the polar boundary layer, the magnetic surfaces focus into
exponential paraboloids (Eq.(\ref{shapeincenterofbl})) as shown in Fig.\ref{fig4}.

\begin{figure}
\epsscale{1}\plotone{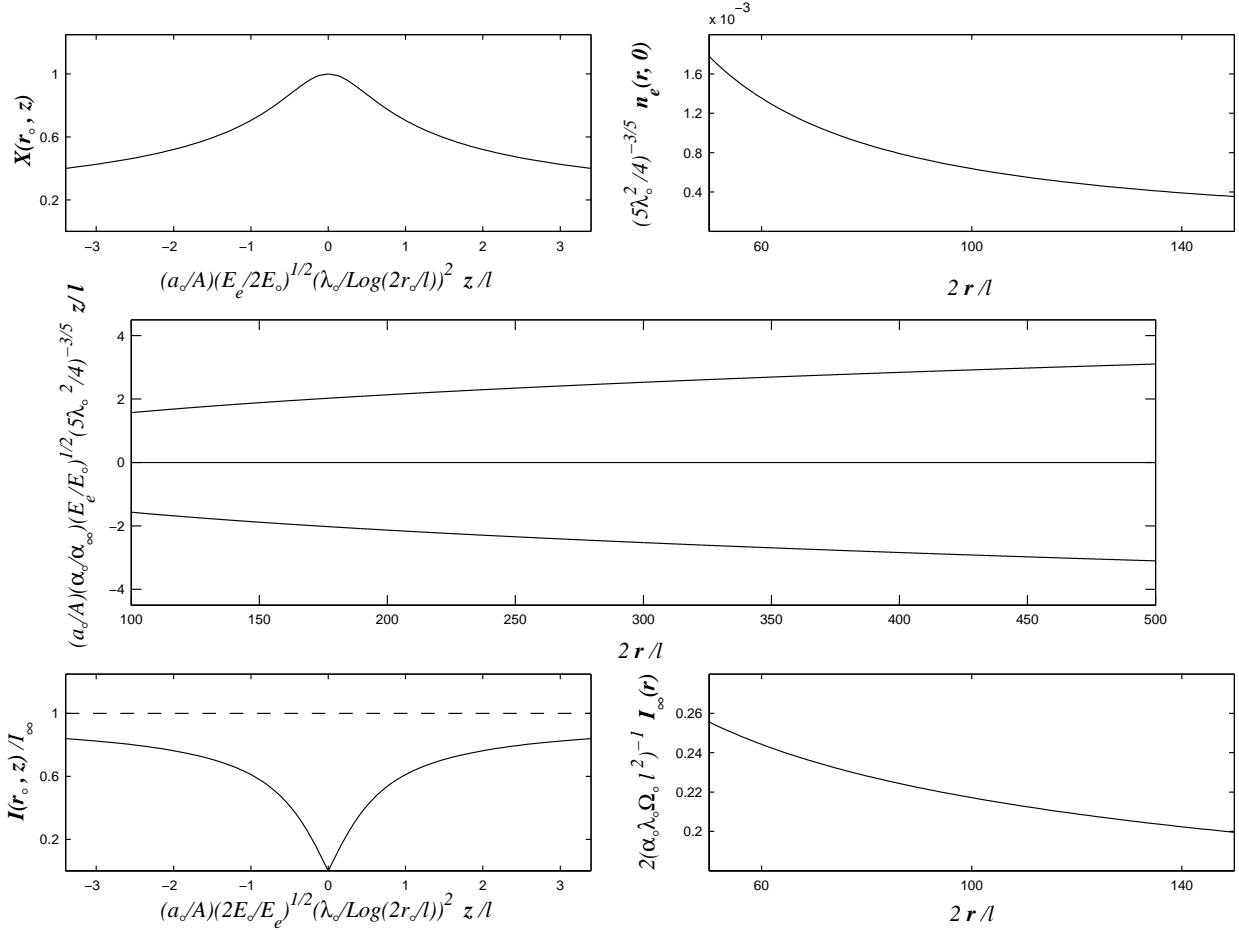}
\caption{The behavior of kinetic winds about the equatorial boundary layer. The panels show:
the magnetic field structure (central); 
the normalized density $n_e$ = $\frac{\rho(r,0)}{\rho_{A0}}$ 
in the equatorial plane vs. distance along the equatorial plane (upper right); the total
current integrated about the equatorial plane vs. distance 
along the equatorial plane (lower right); the ratio of the
density to its equatorial value at the same $r$ across the equatorial sheet (upper left) 
and the integrated current across the equatorial sheet (lower left). 
The latter two quantities are plotted for some arbitrary value of $r$, $r_0$.
For simplicity, $\Omega_e$, $\alpha_e$ and $Q_e$ have been 
set equal to $\Omega_0$, $\alpha_0$ and $Q_0$ respectively. 
The field lines (central frame, Eq.(\ref{zofRinequatBLparab}))
correspond to $\frac{a}{A}$ = 0.9. Their parabolic shape is not 
clearly visible on the scale of the plot. The plots in the right 
panels are from Eqs.(\ref{bennetnull}) and (\ref{assrhonullparab}), while those in the
left panels are from Eqs.(\ref{Xneutralgeneral}), (\ref{alphanullparam}), 
(\ref{Inearnull}) and (\ref{alphascalingatneutral}) 
with $\nu$ = $\frac{1}{2}$. For these plots, the adiabatic 
index $\gamma$ has been taken to be equal to $\frac{4}{3}$.
\label{fig5}}
\end{figure}

\begin{figure}
\epsscale{0.75}\plotone{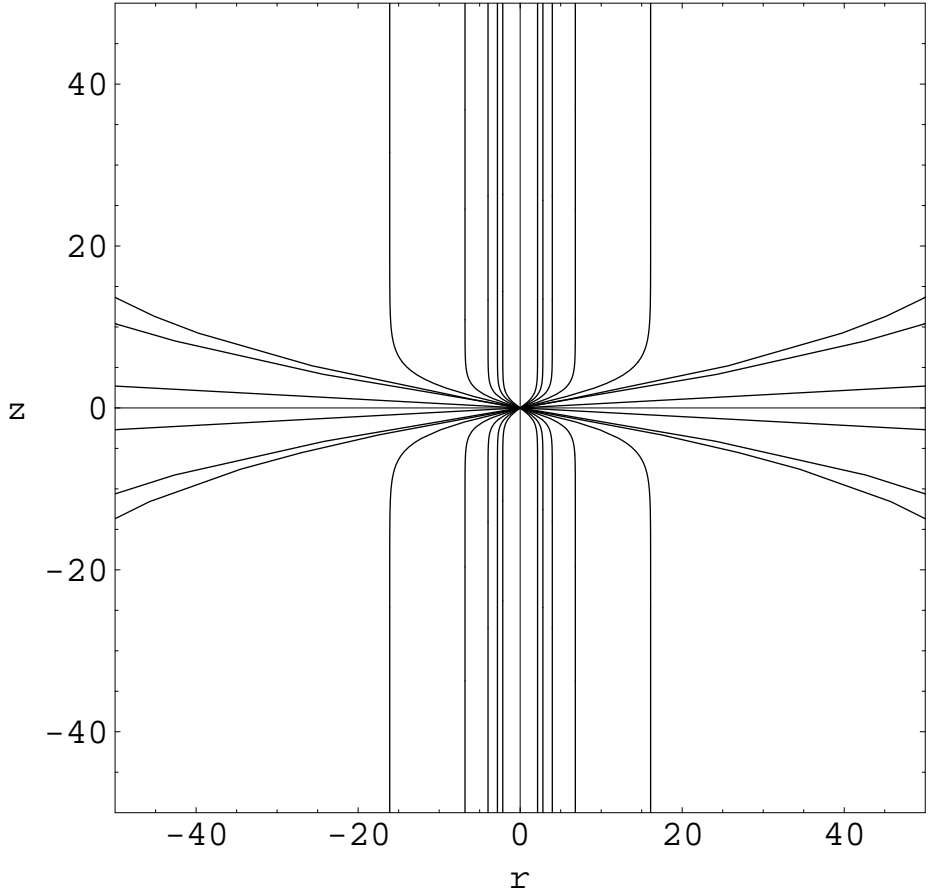}
\caption{The magnetic field structure for Poynting
jets in the field region, from Eqs.(\ref{solpsiofa})
and (\ref{rinftyofa}).
From pole to equator, the contours are for
$(a/A) =$ $0.4$, $0.5$, $0.6$, $0.7$, $0.77$, $0.83$, $0.85$ and $0.95$.
The functions $E(a)$ and $\Omega(a)$ are
constants, while $\alpha(a) = (I \Omega /E)
\lbrace 1 + \left[((a/A) -0.8)^2 /(1 -(a/A))^{1/2} \right] \rbrace $.
The unit of the plot for the variables $r$ and $z$ is the
scale $\ell$ in Eq.(\ref{normalparab}).
An interpolation formula has been used to connect
the asymptotic structure of the field
to a split monopole field near the origin.
Eqs.(\ref{solpsiofa}) and (\ref{rinftyofa})
become increasingly accurate with larger $r/\ell$ and
$z/\ell$. The scale of the transition between the split monopole
structure near the origin and the asymptotic solution has been chosen arbitrarily.
\label{fig6}}
\end{figure}

\noindent
(iii) Near a null surface, which could be the equatorial plane, 
the lines are concave, bending towards the equator deep inside
the neutral boundary layer (Eq.(\ref{zofRinequatBLparab})).  
The magnetic surfaces become
straight lines with a logarithmic
correction (Eq.(\ref{shapeoutskirtequaparab})) at the edge of the
layer. These lines are convex, bending away from the equatorial plane,
outside the neutral sheet as shown in Fig.\ref{fig5}.

\noindent
II. For winds carrying Poynting flux at infinity:

\noindent
(i) In the free field the solutions
asymptote to nested cylindrical and conical magnetic surfaces.
(Eqs.(\ref{rinftyofa}) and (\ref{solpsiofa})) as shown in Fig.\ref{fig6}.

\begin{figure}
\epsscale{1}\plotone{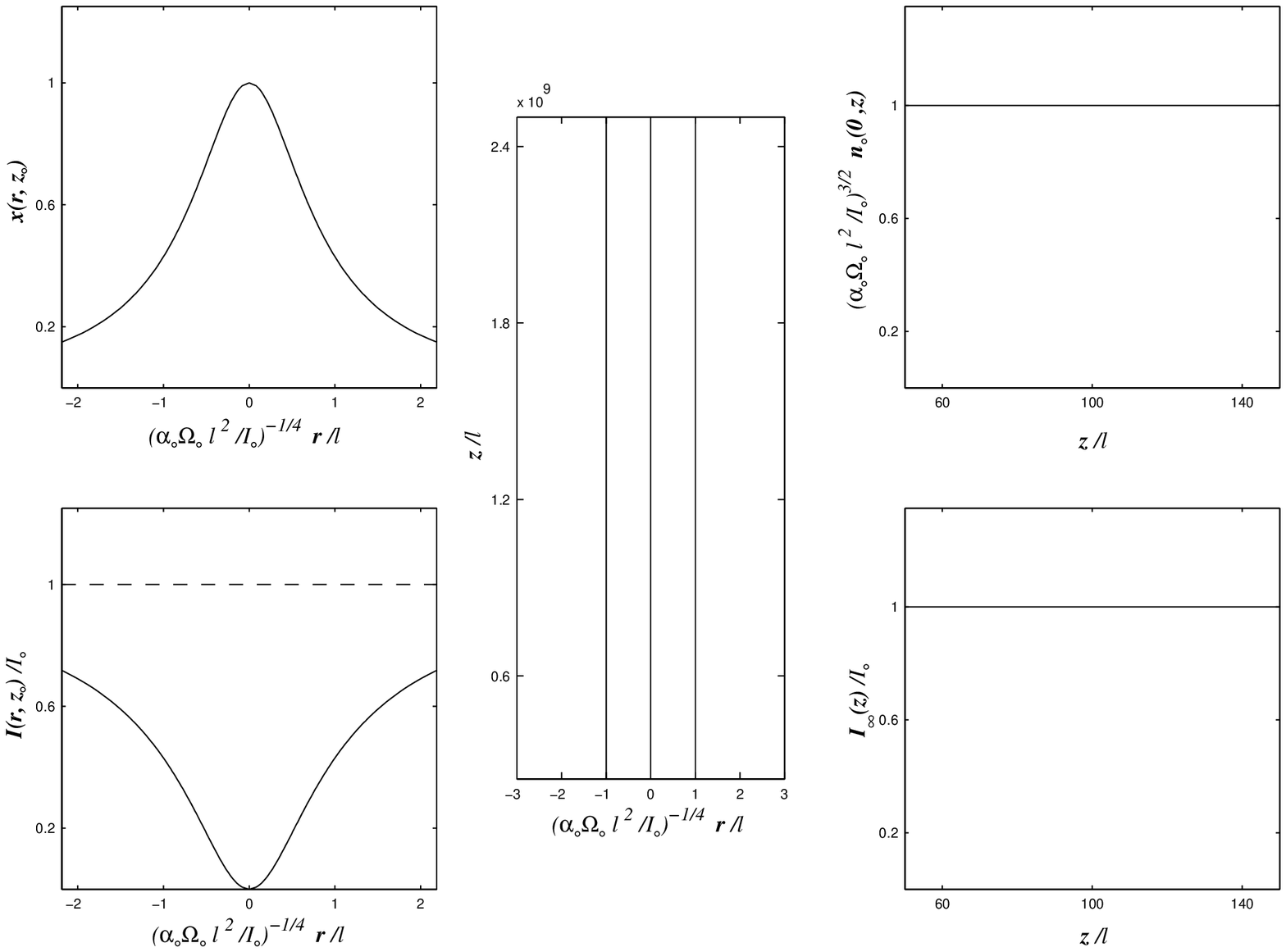}
\caption{The behaviour of Poynting jets near the polar axis. The plots show:
magnetic field structure (central frame); the normalized density at the polar
axis vs. the distance along the polar axis (upper right); the total current
integrated about the polar axis vs. the distance along the polar axis (lower 
right); the ratio of the density to its polar value at the same $z$ across the
polar pinch (upper left) and the integrated current across the polar axis
(lower left). The latter two quantities are plotted for some arbitrarily 
chosen $z_0$. The central frame shows that the field lines are exact cylinders
(Eq.(\ref{radiusofpolarBL})). The right panel curves use 
Eqs.(\ref{bennetparab}) and (\ref{normalparab}). The left panel curves 
are obtained from Eqs.(\ref{defdensratio}), (\ref{parabpsiblofx}), 
(\ref{parabablofx}), (\ref{Iasympt}) and (\ref{normalparab}).
$I_0$ here is the absolute minimum
of $\frac{\alpha E}{\Omega}$. \label{fig7}}
\end{figure}

\noindent
(ii) In the polar boundary layer they are cylindrical 
(Eqs.(\ref{parabpsiblofx})--(\ref{parabablofx})) as shown in Fig.\ref{fig7}.

\begin{figure}
\epsscale{1}\plotone{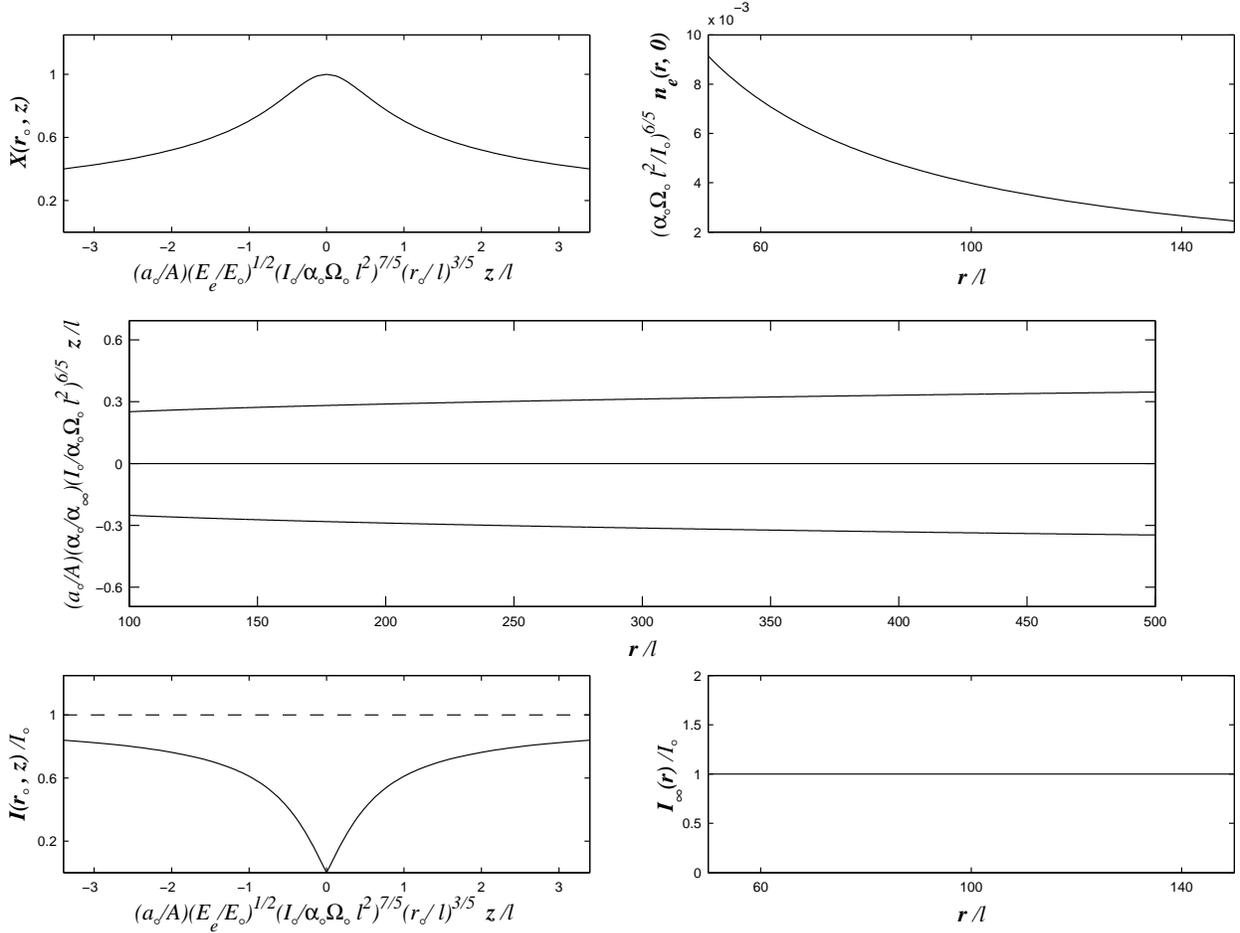}
\caption{The behaviour of Poynting jets about the equatorial boundary layer.
The plots show: magnetic field structure 
(central frame, from Eq.(\ref{zofRinequatBLcyl})); the normalized density
$n_e$ = $\frac{\rho(r,0)}{\rho_{A0}}$ in the equatorial plane vs. distance along
the equatorial plane (upper right); the total current integrated about the 
equatorial plane vs. distance along the equatorial plane (lower right); the 
ratio of the density to its equatorial value at the same $r$ 
across the equatorial sheet (upper left) and the integrated current across the equatorial sheet
(lower left). The latter two quantities are plotted for some arbitrary value of
$r$, $r_0$. The field structure (central frame) is plotted for 
$\frac{a}{A}$ = 0.9. The plots in the right 
panels are from Eqs.(\ref{bennetnull}) and 
(\ref{assrhonullcylind}), while
those in the left panels are from Eqs.(\ref{Xneutralgeneral}), 
(\ref{alphanullparam}), (\ref{Inearnull}) and (\ref{alphascalingatneutral})
with $\nu$ = $\frac{1}{2}$. For these
plots, the adiabatic index $\gamma$ has been taken to be equal to
$\frac{4}{3}$.
 \label{fig8}}
\end{figure}

\noindent
(iii) Near a null surface, which could be the equatorial 
plane, the lines are concave, bending towards the equator deep inside
the neutral boundary layer (Eq.(\ref{zofRinequatBLcyl})).  
The magnetic surfaces become
straight lines (Eq.(\ref{zofRequatoroutskirt})) at the edge of the
layer. These lines remain straight outside the neutral sheet as shown in Fig. \ref{fig8}.

From an observational point of view, the polar and null surface
boundary layers which carry residual electric current may stand out
against the field-regions, both because of their large density
contrast and because they are a source of free energy. This free
energy (associated with the currents) has the potential of making them
active, by the development of instabilities. The density about the
pole does not decline with distance in the case of Poynting jets, or
does it only very slowly in the case of kinetic winds.  It may be that
what is observed as a jet is the dense and active polar boundary layer
of a flow developped on a much larger angular scale. On these larger
scales the flow may be difficult to observe because of its very low
density and current. Null-surface boundary layers, for example
equatorial ones, could be observed in association with the jets,
although their density and activity is expected to decline more
rapidly with distance than on the polar axis, because of geometrical
effects. Their detection could be misinterpreted as accretion disks.

The conclusions reached in this paper are of course only valid under
our assumptions of stationnarity, axisymmetry, polytropic
thermodynamics, perfect MHD and unconfinement. None of those is
expected to be exactly satisfied in reality.  Nevertheless we believe
that the electric circuit picture which emerges from our analysis
should be robust against relaxing a number of these assumptions.
Our description of current closure must remain a feature of the solution, because current
leakage between main regions of current flow should remain weak when
the large regions between them suffer little instability. This is
expected to be the case for regions devoid of current and with only
small velocity gradients. Forces competing with hoop stress must also
remain small under rather general conditions and are probably confined
to spatially limited regions.  By contrast, the distribution of flux
in the different boundary layers would be modified by the
consideration of non-ideal effects, such as turbulence resulting from
the development of non axisymmetric MHD instabilities.  The
stationnarity assumption is valid on a scale smaller than the size of
the cavity carved by the wind in the ambient medium. The wind is bordered
by a time dependent region featuring an expanding shock system.
Actually this region is where the residual electric currents flow from
the circum-polar region to the return regions. That the system, on a
scale less than that of the global cavity, reaches stationarity seems
to be a reasonable assumption.

In the following paper \citep{HNII}
we generalize these results to the relativistic
regime.  An additional paper \citep{HNIII}
discusses whether magnetized outflows are
kinetic-energy-dominated or carry Poynting flux.

\acknowledgements

The authors thank the Space Telescope Science Institute and the Johns
Hopkins University for continued support to their collaboration.  JH
also thanks the EC Platon program (HPRN-CT-2000-00153) and the Platon
collaboration. CN is pleased to thank the Director of ESO for support
and hospitality during which time this paper was completed. We also
thank Sundar Srinivasan for significant help with the figures.

\appendix

\renewcommand{\theequation}{A-\arabic{equation}}
\setcounter{equation}{0}
\section{ Curvature of Poloidal Field Lines}
\label{appreducTFcurvature} 

We write the projection of the equation of motion on the normal to the
magnetic surfaces as (see for example \citet{HeyvNorman89}):
\begin{eqnarray}
{\alpha \over \rho r} \left[
{\partial \over \partial z}
{\alpha \over \rho r} {\partial a \over \partial z}
+ {\partial \over \partial r}
{\alpha \over \rho r} {\partial a \over \partial r} \right]
- {1\over \mu_0 \rho r} \left[
{\partial \over \partial z}
{ 1 \over r} {\partial a \over \partial z}
+ {\partial \over \partial r}
{ 1 \over r} {\partial a \over \partial r} \right] =
E'(a) - {Q'(a) \over \gamma - 1} \rho^{\gamma -1}
\nonumber\\
+ {\alpha'\over \alpha}{\mu_0 \alpha^2 \rho\over r^2}
\left({L-r^2 \Omega \over \mu_0 \alpha^2 - \rho}\right)^2
- {\rho \over r^2} (L' - r^2 \Omega')
\left( {L - r^2 \Omega \over \mu_0 \alpha^2 - \rho}\right)
- { L L' \over r^2} .
\label{transfield} \end{eqnarray}
where primed quantities are  derivatives with respect to $a$ of surface functions.
The transfield equation 
(\ref{transfield}) can be transformed by  making use of the following relations which 
can be derived by explicit calculation of 
$ ({\rm{curl}} \ \vec{v}_P \times \vec{v}_P)$ and
$ ({\rm{curl}} \ \vec{B}_P \times \vec{B}_P)$, using also Eq.(\ref{defalpha}):
\begin{equation}
{\alpha \over \rho r} \left[
{\partial \over \partial z}
{\alpha \over \rho r} {\partial a \over \partial z}
+ {\partial \over \partial r}
{\alpha \over \rho r} {\partial a \over \partial r} \right]
= - {{\vec{\nabla} a}\over{\vert \nabla a \vert^2}}
\left((\vec{v}_P \cdot \vec{\nabla}) \vec{v}_P 
- \vec{\nabla} ({v_P^2 \over 2})\right)
\label{rotvpcrossvp} \end{equation}
\begin{equation}
{1 \over r} \left[
{\partial \over \partial z}
{1 \over r} {\partial a \over \partial z}
+ {\partial \over \partial r}
{1 \over r} {\partial a \over \partial r} \right]
= - {{\vec{\nabla} a}\over{\vert \nabla a \vert^2}}
\left((\vec{B}_P \cdot \vec{\nabla}) \vec{B}_P
- \vec{\nabla} ({B_P^2 \over 2})\right)
\label{rotbpcrossbp} \end{equation}
The vectors $\vec{v}_P$ and $\vec{B}_P$
are separated
in modulus and direction as
$ \vec{v}_P = v_P \ \vec{t}$ and $ \vec{B}_P = B_P \  \vec{t}$, 
with $\vec{t}$ the unit vector tangent to the poloidal field 
line. This gives:
\begin{equation}
(\vec{v}_P \cdot \vec{\nabla}) \vec{v}_P = 
v_P^2 \ (\vec{t} \cdot \vec{\nabla}) \vec{t}
+ \vec{t} \  (\vec{t} \cdot \vec{\nabla}) (v_P^2 /2) 
\label{vpgradvp} \end{equation}
and a similar equation for $\vec{B}_P$.
Then use is made of the Fresnet formula 
$(\vec{t} \cdot \vec{\nabla}) \vec{t} =
 \vec{n} (d \psi / ds)$ 
whith $\psi$ the angle between the vector $\vec{t}$ and its 
projection onto the equatorial plane,  $s$ the curvilinear 
abcissa along the poloidal field line and $\vec{n}$ 
the unit normal vector oriented towards the polar axis
($ \vec{n} = - {\vec{\nabla a} / \vert \nabla a \vert} $).
This transforms equations (\ref{rotvpcrossvp}) and 
(\ref{rotbpcrossbp}) into
\begin{equation}
{\alpha \over \rho r} \left[
{\partial \over \partial z}
{\alpha \over \rho r} {\partial a \over \partial z}
+ {\partial \over \partial r}
{\alpha \over \rho r} {\partial a \over \partial r} \right]
= {v_P^2 \over \vert \nabla a \vert} {d \psi \over ds} 
- {\vec{n} \over \vert \nabla a \vert} \cdot \nabla ({v_P^2 \over 2})
\label{inertiaascurvature} \end{equation}
\begin{equation}
{1 \over r} \left[
{\partial \over \partial z}
{1 \over r} {\partial a \over \partial z}
+ {\partial \over \partial r}
{1 \over r} {\partial a \over \partial r} \right]
= {B_P^2 \over \vert \nabla a \vert} {d \psi \over ds}
- {\vec{n} \over \vert \nabla a \vert} \cdot \nabla ({B_P^2 \over 2})
\label{magtensionascurvature} \end{equation}
The transfield equation (\ref{transfield}) is thus reduced to the form:
\begin{eqnarray} 
{v_P^2 \over \vert \nabla a \vert} (1 - {\rho \over \rho_A})
{d \psi \over ds} = { \vec{n} \cdot \vec{\nabla} \over \vert \nabla a \vert}
\left( {v_P^2 \over 2} \right)
- { 1 \over \mu_0 \rho} { \vec{n} \cdot \vec{\nabla} \over \vert \nabla a \vert}
\left( {B_P^2 \over 2} \right)
\nonumber\\
+ E' - {Q' \rho^{\gamma -1} \over \gamma - 1}
+ {\alpha ' \over \alpha} {\rho_A \rho \over r^2} 
\left({L - r^2 \Omega  \over \rho_A - \rho } \right)^2
- { \rho \over r^2} {(L' - r^2 \Omega ' ) (L -r^2 \Omega) \over \rho_A - \rho} 
-{L L' \over r^2}
\label{trcurvatureform1} \end{eqnarray}
Eliminating $v_P^2 / 2$ on its r.h.s. by Eq.(\ref{Bern}), 
Eq.(\ref{trcurvatureform1}) becomes:
\begin{eqnarray}
{v_P^2 \over \vert \nabla a \vert} (1 - {\rho \over \rho_A})
{d \psi \over ds} =
- { 1 \over \mu_0 \rho} { \vec{n} \cdot \vec{\nabla} \over \vert \nabla a \vert}
\left( {B_P^2 \over 2} \right)
- { 1 \over \rho} { \vec{n} \cdot \vec{\nabla} \over \vert \nabla a \vert}
\left(Q \rho^{\gamma} \right)
-  { \vec{n} \cdot \vec{\nabla} \Phi_G \over \vert \nabla a \vert}
-  { \vec{n} \cdot \vec{\nabla} \over \vert \nabla a \vert} 
\left( {v_{\theta}^2 \over 2}\right)
\nonumber\\
+ { \vec{n} \cdot \vec{\nabla} \over \vert \nabla a \vert}
\left( {r \Omega B_{\theta} \over \mu_0 \alpha}\right)
+ {\alpha ' \over \alpha} {\rho_A \rho \over r^2}
\left({L - r^2 \Omega  \over \rho_A - \rho } \right)^2
- { \rho \over r^2} {(L' - r^2 \Omega ' ) (L -r^2 \Omega) \over \rho_A - \rho}
-{L L' \over r^2}
\label{trcurvatureform2} \end{eqnarray}
We further eliminate $v_{\theta}$ for $L$ and $I$ by using 
Eqs.(\ref{IofBtheta}) and  (\ref{defL}) in 
Eq.(\ref{Btheta}), so that
\begin{equation}
{{L - r^2 \Omega}\over{\rho_A - \rho }} = - {{I}\over{\alpha \rho}}
\label{singratioversusI} \end{equation}
This then yields the following form of the transfield equation
\begin{eqnarray}
{v_P^2 \over \vert \nabla a \vert} (1 - {\rho \over \rho_A})
{d \psi \over ds} =
- { 1 \over \mu_0 \rho} { \vec{n} \cdot \vec{\nabla} \over \vert \nabla a \vert}
\left( {B_P^2 \over 2} \right)
- { 1 \over \rho} { \vec{n} \cdot \vec{\nabla} \over \vert \nabla a \vert}
\left(Q \rho^{\gamma} \right)
-  { \vec{n} \cdot \vec{\nabla} \Phi_G \over \vert \nabla a \vert}
\nonumber\\
+ {{(L - I / \alpha)^2}\over{r^3}} { \vec{n} \cdot \vec{\nabla}r \over \vert \nabla a \vert}
- {1 \over 2 r^2}  { \vec{n} \cdot \vec{\nabla}r \over \vert \nabla a \vert}
\left( {I^2 \over \alpha^2}\right)
+ {1 \over r^2} (L - r^2 \Omega) { \vec{n} \cdot \vec{\nabla}r \over \vert \nabla a \vert}
\left( {I \over \alpha}\right)
+ {\alpha' \over \alpha} {\mu_0 I^2 \over \rho r^2}
\label{trcurvatureform3} \end{eqnarray}
Eliminating  $\rho r^2$ for  $I$ 
by solving Eq.(\ref{singratioversusI}), 
Eq.(\ref{trcurvatureform3}) becomes:
\begin{eqnarray}
v_P^2 (1 - {\rho \over \rho_A}) {d \psi \over ds} =
{1 \over \rho} { \vec{\nabla} a \cdot \vec{\nabla} \over \vert \nabla a \vert}
\left({(\nabla a)^2 \over 2 \mu_0 r^2} \right)
+ {1 \over \rho} { \vec{\nabla} a \cdot \vec{\nabla} \over \vert \nabla a \vert}
\left( Q \rho^{\gamma}) \right)
\nonumber\\
+ { \vec{\nabla} a \cdot \vec{\nabla} \Phi_G \over \vert \nabla a \vert}
- {{(L - I / \alpha)^2}\over{r^3}} {{\partial a /\partial r}\over{\vert \nabla a \vert}}
+ {\mu_0 I \over \rho r^2} { \vec{\nabla} a \cdot \vec{\nabla} I \over \vert \nabla a \vert}
\label{trcurvaturewithI} \end{eqnarray}
The centrifugal term, proportional to $\partial a /\partial r$,
simplifies by using Eq.(\ref{singratioversusI}).
In the limits $r \rightarrow \infty$, 
$\rho /\rho_A \rightarrow 0$, $\Phi_G \rightarrow 0$, $v_{\theta} \rightarrow 0$
(as $1/r$), Eq.(\ref{trcurvaturewithI}) 
reduces, by expressing
$ \lim (\rho r^2)$ as $\mu_0 \alpha I / \Omega$, to:
\begin{equation}
v_P^2 {d \psi \over ds} =
{r^2 \Omega \over \mu_0 \alpha I} { \vec{\nabla} a \cdot \vec{\nabla} \over \vert \nabla a \vert}\left( {(\nabla a)^2 \over 2 \mu_0 r^2} +  Q \rho^{\gamma}) \right)
+ {\Omega \over \alpha}  { \vec{\nabla} a \cdot \vec{\nabla} I \over \vert \nabla a \vert}
\label{trfieldassincurvform} \end{equation}
The terms left in this equation are 
still not of the same order of magnitude. This 
is discussed in section \ref{secTFasympt}.

\renewcommand{\theequation}{B-\arabic{equation}}
\setcounter{equation}{0}
\section{Equivalence of Eikonal Equation with Snell's Law}
\label{appsnell}

Let $\vec{n}$ be the unit vector in the direction of $\vec{\nabla} S$,
so that the Eikonal equation (\ref{HamJacobS}) is:
\begin{equation}
\vec{\nabla}S =
N \  \vec{n}
\label{eikonaleq} \end{equation}
The angle of incidence $i$ which appears in Snell's law is the angle between
$\vec{n}$ and
$\vec{\nabla} N$. Taking the curl of equation (\ref{eikonaleq}) gives
$ \vec{\nabla} \times \vec{n} =
\vec{n}  \times \vec{\nabla} N / N$ and the
projection of this equation onto 
the plane perpendicular to $\vec{n}$ gives
\begin{equation}
( \vec{\nabla} \times \vec{n}) \times \vec{n} =
{{\vec{\nabla}N}\over{N}}
- \vec{n} \ {{\vec{n}.\vec{\nabla}N}\over{N}}
= {{\vec{\nabla}_{\perp}N}\over{N}}
\label{gradNperp} \end{equation}
where the subscript $\perp$ indicates the component of $\vec{\nabla} N$
perpendicular to the local vector $\vec{n}$.
Since $\vec{n}$ is a unit vector, 
\begin{equation}
0 = \vec{\nabla} (\vec{n}.\vec{n}) = 
2 (\vec{n}. \vec{\nabla})\vec{n} +
2 \vec{n} \times \left(\vec{\nabla} \times \vec{n}\right)
\label{gradnn}
\end{equation}
This allows to transform 
 the left hand side of Eq.(\ref{gradNperp}) into
$(\vec{n}. \vec{\nabla}) \vec{n}$. Introducing
the curvilinear abcissa $\sigma$ 
along an orthogonal trajectory to
lines $S(r,z)=$ constant in the meridional plane, 
$(\vec{n}. \vec{\nabla}) \vec{n} \equiv d\vec{n}/d\sigma$. Thus 
Eq.(\ref{gradNperp}) is:
\begin{equation}
{{d \vec{n}}\over{d \sigma}}
={{\vec{\nabla}_{\perp} N}\over{N}}
\label{dnoverdsigma} \end{equation}
Since $\vec{n}$ is a unit vector, $d\vec{n}/d\sigma$ 
is perpendicular to $\vec{n}$. According to Eq.(\ref{dnoverdsigma}), 
it is in the incidence plane
defined by the vectors $\vec{n}$ and $\vec{\nabla}N$.
Let $\vec{t}$ be the unit vector in
the plane of incidence which is perpendicular to
$\vec{n}$ and is oriented along the direction of
${{\vec{\nabla}_{\perp} N}/{N}} $.
The projection of Eq.(\ref{dnoverdsigma}) on the 
vector $\vec{t}$ gives:
\begin{equation}
\vec{t} \ . \  {{d \vec{n}}\over{d \sigma}} = -  {{di}\over{d \sigma}}=
\sin i \  \vert \nabla (\ln N) \vert.
\label{tdnoverdsigma} \end{equation}
The minus sign in the second term of (\ref{tdnoverdsigma})
arises  because the curvature of the ray towards
the direction of $\vec{\nabla} N$ causes $i$ to decrease.
Since $\vec{\nabla} \ln N$ is at an angle $i$ to $\vec{n}$,
\begin{equation}
{{d\ln N}\over{d \sigma}}=
\vert \nabla \ln N \vert \  \cos i
\label{presnell} \end{equation}
and Eq.(\ref{tdnoverdsigma}) finally reduces to
\begin{equation}
{{di}\over{d \sigma}}=
- {{\sin i}\over{\cos i}} \  {{d \ln N}\over{d \sigma}}
\label{snelldifferential} \end{equation}
It integrates into Snell's law ($\ln \left(N \sin i\right)=$ constant
following a ray).

\renewcommand{\theequation}{C-\arabic{equation}}
\setcounter{equation}{0}
\section{Field Regions of Winds with a Finite Flux}
\label{appnoparaboloidswithcurrent}

If $I_{\infty}$ is non-zero, the magnetic structure in the asymptotic 
domain consists of a cylindrically collimated core
surrounded by a magnetic structure which encloses poloidal current but
does not carry it. This structure can be of a conical geometry but it
could conceivably also be of any current-enclosing parabolic geometry,
as for example paraboloids of a
constant power exponent \citep{begli94}. Our results
of section \ref{subsecraytracing} show that when this magnetic structure
is constrained to smoothly join the equatorial plane, 
the solution for flaring magnetic surfaces  
consists of nested cones. 
We show here more directly that a conical distribution
is the only possibility for a current-carrying wind blown by 
a  wind source subtending a finite flux.
Paraboloids of a constant exponent
are represented as:
\begin{equation}
z = K(a) r^q
\label{constantqparab} \end{equation}
with  $q$ a constant. The
case $q=1$ corresponds to cones, while
$q$ strictly larger than unity, but constant, corresponds
to current-enclosing paraboloids. Differentiating Eq.(\ref{constantqparab}) gives
\begin{equation}
r \vert \nabla a \vert = {K \over K'} \sqrt{q^2 + r^{2(1-q)} / K^2}
\label{rgradaparabconstantq} 
\end{equation}
Substituting Eq.(\ref{rgradaparabconstantq}) in Eq.(\ref{Iassgeneral}) 
with supposedly constant $I_{\infty}$ gives
\begin{equation}
{K' \over K \sqrt{q^2 + r^{2(1-q)} / K^2}} =
{{\Omega (a) }
\over
{ \mu_0  I_{\infty} \sqrt{2} \sqrt{E(a) - I_{\infty} \Omega (a) /\alpha (a)}}}
\end{equation}
In the large-$r$ limit
the l.h.s reduces, when $q$ is strictly larger than unity (paraboloids), 
to a logarithmic derivative.
It is therefore impossible in this case to meet the condition that
$K=0$ at the equator, the magnetic surface  $a=A$,
if $A$ is finite. This conclusion holds also for more complicated
current-containing structures of parabolic geometry,
such as for example $ z = K(a) e^{qr}$, with constant $q$.
In the case of conical asymptotics,
we recover the results of \citet{HeyvNorman89}.

\renewcommand{\theequation}{D-\arabic{equation}}
\setcounter{equation}{0}
\section{Particular Solutions to the Hamilton-Jacobi Equation}
\label{apppartsolHamJacob}

In this appendix we obtain particular, separable, solutions of the
Eikonal equation (\ref{HamJacobS}) $r \vert \nabla S \vert = 1$.
Squaring it and passing to spherical coordinates,
spherical radius $R$ and latitude $\lambda$,
it can be written as
\begin{equation}
R^2 \left({{\partial S}\over{\partial R}}\right)^2
+\left(\left({{\partial S}\over{\partial \lambda}} \right)^2  
- {{1}\over{\cos^2\lambda}}\right) = 0
\label{HamJacspher} \end{equation}
Separable solutions are of the form
$ S(R, \lambda) = F(R) + G(\lambda)$.
The number $m$ being a real constant,
positive or negative,
the variable separation gives for $F$ and $G$ the
differential equations
\begin{equation}
{{dF}\over{d \ln R}} = m
\label{HJsepR} \end{equation}
\begin{equation}
\left({{dG}\over{d\lambda}}\right)^2 -  {{1}\over{\cos^2\lambda}}  = - m^2
\label{HJseplambda} \end{equation}
which can be integrated into
\begin{equation}
F = m \ \ln R  + C_1
\label{solHJsepR} \end{equation}
\begin{equation}
G = \pm \int_{\lambda_0}^{\lambda} {{d \lambda' \sqrt{1 - m^2 
\cos^2\lambda'}}\over{\cos\lambda'}}  + C_
2
\label{solHJseplambda} \end{equation}
Simple particular solutions can be found.
For  $m=0$, $S$ depends on $\lambda$ alone,
which means that flux surfaces which reach $r = \infty$
asymptote to cones.
For $m= \pm 1$, we have two types of
possible solutions with
$F = \pm \ln R + C_1 $
and $G = \pm \ln(\cos \lambda) + C_2$, the two $\pm$
signs being independent of eachother.
The only physically acceptable solution is
\begin{equation}
S = \pm \ \ln(R \cos \lambda) + C
\label{solHJsepm1} \end{equation}
which gives poloidal field lines (of constant S)
parallel to the polar axis.
The other sign combination must be rejected
since it gives rise to magnetic surfaces
that do not reach $z = \infty$.
Equation (\ref{solHJseplambda}) does not integrate simply in the general
case. The solution can be reduced to quadratures and written as:
\begin{equation}
S = C + m \ \ln R \
\pm \int_0^{\lambda}
{{d \lambda' \sqrt{1 - m^2 \cos^2\lambda'}}\over{\cos\lambda'}}
\label{solHJsep} \end{equation}
For a physical solution, $R$ must be able to
approach infinity at constant $a$, that is at constant S. This
is possible only when the sign of the functions of $R$
and $\lambda$ on the r.h.s.
of Eq.(\ref{solHJsep}) are different, and the integral
in $\lambda$ diverges as $R$ grows very large.
For small $ \phi = \pi/2 - \lambda$,
the angular integral diverges as 
$\pm \ln \phi$, and the poloidal field lines approximately become lines
\begin{equation}
R^{\vert m\vert} = {C'(S) \over \phi}
\label{solHJsepsmallphi} \end{equation}
Since $R \approx z$ and $\phi \approx (r/z)$, this gives
the cylindrical radius $r$ in terms of $z$ as:
\begin{equation}
r =  C'(S) z^{(1- \vert m\vert)}
\label{solHJsepquasivert}\end{equation}
The physically consistent solutions correspond to $\vert m \vert \le 1$,
since  otherwise $r $ would not grow large with $z$,
contrary to assumptions made to derive this
 approximate form of the transfield equation.
The solution consists in this case of magnetic
surfaces of an asymptotically parabolic shape,
with a power-law exponent independant of the surface.
The particular cases $\vert m \vert = 1$
and $m = 0$ are of course recovered in this more general family of solutions.
Note however that paraboloids with constant exponent
are not valid solutions for a  wind source subtending a finite
flux (Appendix \ref{appnoparaboloidswithcurrent}). 
Therefore, in this class of separable solutions,
only those which are asymptotically conical
with a cylindrical core of poloidal current-carrying magnetic surfaces
are acceptable.

\renewcommand{\theequation}{E-\arabic{equation}}
\setcounter{equation}{0}
\section{Matching across Null Surface Boundary Layers}
\label{appcells}

We consider here current-carrying jets with several null magnetic surfaces.
The function $I_{\infty}(b)$ is non zero, but it has sign jumps at the
crossing of the boundary layers about null surfaces, which in general
are located between pole and equator.  The equator itself is, 
by symmetry, a magnetic surface, null or not.
Null surfaces separate the space into a number of cells.  The
equatorial plane is embedded in or is bordering one of them.  Consider
these different cells in turn, beginning with one containing the
equator, or being contiguous to it. From our results of section
\ref{subsecraytracing}, the orthogonal trajectories to magnetic
surfaces (rays), are circles centered on the polar axis.  They must be
perpendicular to the equator, which is a particular magnetic surface,
and so, they are in this region (asymptotically) centered on the
origin. The null surface which borders the equatorial cell polewards,
being perpendicular to these rays, must be a cone centered on the
origin. This latter surface is then also perpendicular to the rays
approaching it from the cell polewards, in which $I_{\infty}$ assumes
a value which differs by its sign from the one in the equatorial cell.
The rays in this more poleward cell are also circles centered on the
polar axis. They  must in fact, from this boundary condition, be
centered at the origin.  It is easy to extend this reasoning
recurently to show that the property of the rays to be centered at the
origin passes from the equatorial cell to all the other cells.  This
completes the proof that the magnetic surfaces which reach to
infinitely large values of $r$ in a Poynting jet must asymptotically
be conical. Let us again repeat that such a structure must enclose a
certain flux about the polar axis in which this finite poloidal
current is actually flowing, and where poloidal field lines {\it{do
not}} follow to infinite values of $r$.

\renewcommand{\theequation}{F-\arabic{equation}}
\setcounter{equation}{0}
\section{Flux in the Polar Boundary Layer }
\label{appfluxinbl}

The approximate solution which we have deduced in section \ref{secpolarbl}
has been obtained
assuming the first integrals $E$, $\Omega$, $Q$ to be almost constant in
the polar boundary layer.
The validity of this  can be judged of by calculating
the flux trapped in the polar boundary layer,
given by equation (\ref{parabablofx}) for, say,
$x \approx 1/e$.  This identifies, using Eq.(\ref{bennetparab}), 
the flux variable at the edge of the boundary layer to be:
\begin{equation}
a_{bl} \approx  {1\over\sqrt{2}}\
{\mu_0 I_{\infty}\over\Omega_0}\
\sqrt{E_0 - {I_{\infty} \Omega_0\over\alpha_0}}
\label{fluxinblofI} \end{equation}
For $E$, $\Omega$, $Q$ to be indeed almost constant in the polar 
layer, $a_{bl}$ should  be significantly less than the total
flux $A$. Note that $I\Omega/\alpha$ is of order of the specific
energy associated to the Poynting flux $E_p$.  Then we find
\begin{equation}
{a_{bl}\over A} \ \approx \
{{\mu_0 \alpha E_p v_{\infty}}\over{\Omega^2 A}}\ \approx \
{{\rho_A E_p v_{\infty}}\over{\Omega^2 \dot{M}}} \ \approx \
{{E_p v_{\infty}}\over{\Omega^2 R^2_A v_{PA}}} \ \approx \
{E_p\over E}\
{E \over\Omega^2 R_A^2}\
{{v_{\infty}}\over{v_{PA}}}
\label{estimatefluxinbl} \end{equation}
For fast rotators, the asymptotic velocity $v_{\infty}$ is not
much larger than $v_{PA}$, and $E$ is of order $\Omega^2 R_A^2$. Then
the boundary layer flux can only be  a small amount of the total flux
for flows which are largely, but not totally,
kinetic-energy-dominated at infinity.

By contrast, flows which bring a significant
Poynting energy flux to infinity would
have a large part of their asymptotic flux trapped in
a high pressure zone of cylindrical geometry. Our approximate solution
(\ref{parabablofx})--(\ref{parabpsiblofx})
which assumed little variation of the constants of the motion
with flux over the polar boundary layer 
becomes at best  schematic in this case.
When so, the cylindrical pinch equations should be solved exactly in the
circumpolar region where pressure is important. So would it also be
if $(r_{\infty}/r_A)$ would not be large, as we assumed, in a
substancial fraction
of the polar boundary layer:  other forces than gas pressure
would then have to be taken into account, for example
the centrifugal force. 
The circum-polar structure  should
then be solved for numerically,
in the framework of some specific model
where the first-integrals would be explicitly prescribed.
This numerically-determined solution would still behave for large $r$'s as
Eq.(\ref{solpsiofa}).

\renewcommand{\theequation}{G-\arabic{equation}}
\setcounter{equation}{0}
\section{Mass-to-Magnetic Flux Ratio at a Null Surface}
\label{appalphaatnullsurf}

Near a null magnetic surface of flux variable $a = a_n$
the quantity  $\alpha$ generally diverges as
$\vert a_n -a \vert^{-1/2}$, or, more exceptionnally,
as some negative power of $\vert a_n -a \vert$,
with exponent smaller in absolute value than unity.
This can be seen from Eq.(\ref{defalpha}). 
The quantity  $\alpha$, being a first-integral,
can be evaluated anywhere, for example at a reference sphere of radius $R_i$
close to the wind source. 
If there is to be wind flowing on the surface $a_n$, 
the mass flux $\rho v_P$
should not vanish, while $B_P$ vanishes  as $a \rightarrow a_n$. 
Let $\mu(a)$ be the colatitude  at
which the magnetic surface $a$ is rooted on this
reference sphere and let us note the angle $\mu(a_n)$ as $\mu_n$. 
Since, close to the source, the
magnetic field is essentially undistorted with respect to  
the potential field created by the object
at the source of the wind, 
$B_P$ is expected to vanish, at given $R= R_i$, proportionally to 
$\vert \mu (a_n) - \mu(a) \vert^n$ as $\mu$ approaches $\mu_n$.
The exponent $n$ is usually equal to unity and perhaps may sometimes be
larger. 
The flux is the surface integral
of the component of the field normal to the reference sphere. So we get,
for $\mu(a) \le \mu_n$ and $a \le a_n$, say:
\begin{equation}
(a_n - a) \ \approx \ \int_{\mu(a)}^{\mu_n}
2 \pi R_i^2 \  \sin \mu_n \ d \mu' \  (\mu_n - \mu')^n \ 
=  {{2 \pi R_i^2 \sin \mu_n } \over {n+1}}
\  (\mu_n - \mu(a))^{n + 1}
\label{a*minusaintermsofdtheta} \end{equation}
For $B_P \sim \vert  \mu_n - \mu(a) \vert^n$ this gives
\begin{equation}
B_P \approx   (a_n - a)^{n/(n + 1)}
\label{Bpneara*intermsofdtheta} \end{equation}
Thus $\alpha(a)$ ($= \rho v_P/B_P$) diverges as
$({a_n - a})^{- n / (n+1)}$ when $a \rightarrow a_n$. Since $n$ is
expected to usually be equal to unity 
this implies, as expressed by equation (\ref{divalpha}), that 
\begin{equation}
\alpha(a) \approx \vert ({a_n - a}) \vert^{-1/2}
\label{divalphaappendix} \end{equation}
More generally, with $\nu = (n/n+1) <1$, 
\begin{equation}
\alpha(a) \approx \vert ({a_n - a}) \vert^{\nu}
\end{equation}

\renewcommand{\theequation}{H-\arabic{equation}}
\setcounter{equation}{0}
\section{Magnetic Surfaces in Field Regions of Kinetic Winds}
\label{appgeneralparaboloids}

The solution in the field-region of kinetic winds
is not given accurately enough by (\ref{paraboloidshape}) 
at non-paraxial surfaces bordering the equator 
in the field-region. In this appendix we calculate their shape
which is to be found  
by integrating equations (\ref{fieldlineeq})
without any further approximation.
From Eq.(\ref{solpsiofa}),
using Eq.(\ref{chiofa}) and taking the 
large-$\chi$ limit 
we obtain:
\begin{equation}
\cos\psi = {{2}\over
                {\left({{2 b}\over{\ell}}\right)^{k(a)} +
                \left({{2 b}\over{\ell}}\right)^{-k(a)}}}
\label{cospsiofaandRgeneral} \end{equation}
\begin{equation}
\sin\psi = {{\left( {{2 b}\over{\ell}} \right)^{k(a)} -
                \left( {{2 b}\over{\ell}} \right)^{-k(a)}}\over
                {\left( {{2 b}\over{\ell}} \right)^{k(a)} +
                \left( {{2 b}\over{\ell}} \right)^{-k(a)}}}
\label{sinpsiofaandRgeneral} \end{equation}
where  $k$ is  defined by Eq.(\ref{cospsiexponent}). The coordinates 
$r$ and $z$ are then given in terms of the spherical distance $b$
by the differential equations
\begin{equation}
{{ dz}\over{\ell}} = {{\left( {{2 b}\over{\ell}} \right)^{k} -
                \left( {{2 b}\over{\ell}} \right)^{-k}}\over
                {\left( {{2 b}\over{\ell}} \right)^{k} +
                \left( {{2 b}\over{\ell}} \right)^{-k}}}\
                {{ db}\over{\ell}}
\label{diffeqzofRgeneral} \end{equation}
\begin{equation}
{{dr}\over{\ell}} = {{2}\over
                {\left({{2 b}\over{\ell}}\right)^{k} +
                \left({{2 b}\over{\ell}}\right)^{-k}}}\
                {{db}\over{\ell}}
\label{diffeqrofRgeneral} \end{equation}
Eqs.(\ref{diffeqzofRgeneral}) (\ref{diffeqrofRgeneral}) 
can be reduced to quadratures to give the 
following  parametric representation of
these surfaces, taking ${{b}\over{\ell}} = \lambda$ as
the parameter: 
\begin{equation}
{{z}\over{\ell}} = \lambda  -  \int_0^{2 \lambda}\
{{du}\over{u^{2k}+1}}
\label{zofRgeneral} \end{equation}
\begin{equation}
{{r}\over{\ell}} = {{1}\over{k+1}} \  \int_0^{\left( 2\lambda \right)^{k+1}}
{{du}\over{u^{(2k/(k+1))}+1}}
\label{rofRgeneral} \end{equation}
Again, $k$ is defined by Eq.(\ref{cospsiexponent}).
Let us  calculate the radius of curvature of these 
field lines as a function of the parameter $\lambda$. 
The line element along them is found to be $ds = d \lambda$. 
The unit tangent to the line has components
\begin{equation}
t_r = {{dr}\over{ds}} = {{ 2 \ (2 \lambda)^k}\over{(2 \lambda)^{2k} +1}}
\label{unittangentr} \end{equation}
\begin{equation}
t_z = {{dz}\over{ds}} = {{ (2 \lambda)^{2k} - 1}\over{(2 \lambda)^{2k} +1}}
\label{unittangentz} \end{equation}
and the vector $d \vec{t} /ds$ is then easily calculated 
to be equal to $\vec{n}/R_c$, where $\vec{n}$ is a unit vector perpendicular
to $ \vec{t}$ and the curvature radius is
\begin{equation}
R_c = {{ (2 \lambda)^{2k} +1}\over{ 4 k \ (2 \lambda)^{k -1} }}
\label{curvfieldlinesexplicit}
\end{equation}
Depending on the field line, the parameter $k$ 
varies between zero near the equator and unity at the pole. At 
large distances  $r$ scales as $ (2 \lambda)^{1-k}/(1-k)$
while $R_c$ scales as $(2 \lambda)^{1+k}/4 k$. 

\renewcommand{\theequation}{I-\arabic{equation}}
\setcounter{equation}{0}
\section{Exit Angle from the Equatorial Boundary Layer}
\label{appangleofexitfromeqbl}

Let $X_{ex}$ be the value of the parameter $X$ which corresponds to
the outer edge of the equatorial boundary layer. $X_{ex}$ may be taken as equal to
$1/2$ or $1/e$, say. Assuming $\alpha(a)$ 
to be well approximated in this region by Eq.(\ref{divalpha}) with $\nu = 1/2$,
the flux parameter $a_{ex}(b)$ of the magnetic surface which exits
the boundary layer at $b$
is given by
\begin{equation}
\left(A - a_{ex}(b)\right) \approx
\left({{2 Q_e \mu_0 \lambda^2}\over{ \Omega_e^2 X_{exit}^2}}\right)
\left({{2 Q_e}\over{ \mu_0 J_m^2 }}\right)^{ {{2 - \gamma}\over{\gamma}} }
{{ \left( \ln (2 b/\ell) \right) ^{ {{2 m (2 - \gamma)}\over{\gamma}} }  }\over{
b^{ {{ 4 (\gamma  - 1)}\over{\gamma}} }   }}
\label{aexit} \end{equation}
Incidently this shows that the residual flux in the
equatorial boundary layer declines with distance $b$.
The partial derivative $(\partial z/\partial b)_a$ calculated at $a = a_{ex}(b)$ is
the slope of the poloidal field line
which exits the equatorial boundary layer at $b$. 
It is given in terms of $b$ and $z$ by
\begin{equation}
\left({{\partial z}\over{\partial b}}\right)_a
= {{2 - \gamma}\over{\gamma}} \left({{ z}\over{b }}\right)_{a_{ex}}
\label{slopeandzoverR} \end{equation}
Using Eq.(\ref{aexit}) and 
Eqs.(\ref{behaverhoequat})-(\ref{zofRequatoroutskirt}), 
the following expression is obtained:
\begin{equation}
\left( {{\partial z}\over{\partial b}}\right)_{a_{ex}(b)} = \left({{2
- \gamma}\over{\gamma}} \right) {{2 \lambda^2}\over{\sqrt{2 E_e} }}
\sqrt{{{2 Q_e \mu_0 }\over{ \Omega_e^2 X_{exit}^2 }} } \left({{2 Q_e
}\over{\mu_0 J_m^2}} \right)^{{{4 -\gamma}\over{2 \gamma}} } {{ \left(
( \ln (2 b/\ell) )^m \right)^{{{4 -\gamma}\over{\gamma}} } }\over {
b^{{{4(\gamma -1)}\over{\gamma}} } }}
\label{slopeatexit} \end{equation}
This slope approaches zero as $b$ approaches infinity, as needed for
consistency of our analysis in section \ref{subsecraytracing}, 
in particular for supporting the idea
that orthogonal trajectories to
magnetic surfaces become closer 
and closer to centered spheres as their radius increases.

\renewcommand{\theequation}{J-\arabic{equation}}
\setcounter{equation}{0}
\section{Asymptotic Ordering of Lorentz Forces}
\label{appcomponentsofLorentz}

All components of $\vec{B}$  and $\vec{j}$
approach zero at large distances, though not at the same rate.
In the field-region, $\vec{B}_P$, given by Eq.(\ref{Bpversusa}),
declines with spherical distance as $1/b^2$, while 
$\vert \vec{B}_{\theta} \vert = \mu_0 I_{\infty} / r$  
may decline slower than  $1/b$, depending on the 
shape of magnetic surfaces. The toroidal current density is:
\begin{equation}
\vec{j}_{\theta} = { 1 \over \mu_0} \ \vec{\nabla} \times \vec{B}_P =
- { 1 \over \mu_0} \ \left( {{\partial}\over{\partial z}} {1 \over r}
{{\partial a }\over{\partial z}} + {{\partial}\over{\partial r}} {1 \over r}
{{\partial a }\over{\partial r}} \right) 
\label{jtheta}
\end{equation}
It declines as $1/r^3$, which may be slower than
$1/b^3$ depending on the geometry of magnetic surfaces. The poloidal
current density $\vec{j}_P$ has a component $\vec{j}_{P \perp}$
orthogonal to magnetic surfaces and a component $\vec{j}_{P \|}$ along
them.  The latter is constrained by Eq.(\ref{solvability}) to vanish,
keeping the same approximation (or ordering) at which
Eq.(\ref{solvability}) itself is valid.  The discussion of
section (\ref{dominantforces}) has shown that the gradient of gas
pressure causes $\vec{j}_{P \|}$ to deviate from zero, such that
\begin{equation}
j_{P \|} B_{\theta} \approx \vert \vec{\nabla} Q \rho^{\gamma} \vert
\sim {Q \over r}  \left({{\alpha a }\over{ b^2 v_{\infty} }}\right)^{\gamma}
\label{jpparallel}
\end{equation}
The component $\vec{j}_{P \perp}$ is approximately:
\begin{equation}
j_{P \perp} \approx { 1 \over b} {{dI_{\infty} }\over{ db}} 
\label{jpperp}
\end{equation}
The magnitude of $j_{P \perp}$ depends on how $I_{\infty}(b)$ converges
to its limit. If this limit is zero, $I_{\infty}$ is given by
Eqs.(\ref{bennetparab})--(\ref{nzeroofRsimple}).  If the limit is the
minimum $I_{sup}$ of the function $(\alpha E/\Omega)$, an expansion of
Eq.(\ref{facteur}) for $I_{\infty}$ close to $I_{sup}$ can be made
resulting in
\begin{equation}
\lambda(b) \sim \ln \left( {{1}\over{ \sqrt{ 1 - {{I}\over{I_{sup}}} }  }} \right)
\label{lambdanearIsup} \end{equation}
When substituted in Eq.(\ref{nzeroofRsimple}) it is found that
$I_{\infty}$ is given by
\begin{equation}
I_{\infty} = I_{sup} \left( 1 - \left({{\ell}\over{b}} \right)^{\xi} 
\right) 
\label{IinftynearIsup} \end{equation}
where $\xi$ is usually a small exponent.
It is then deduced that 
for kinetic winds
\begin{equation}
j_{P \perp} \approx {{Q_0 \alpha_0 \rho_{A0}^{\gamma -1} }\over{ \Omega_0}} 
{{1}\over{b^2 (\ln(b))^2 }} 
\label{jperpforkinetic}
\end{equation}
while for Poynting jets
\begin{equation}
j_{P \perp} \approx I_{sup} {{\xi}\over{b^2}}
\left( {{\ell}\over{ b}} \right)^{\xi + 2}
\label{jperpforpoynting}
\end{equation}
Eqs.(\ref{jperpforkinetic}) and (\ref{jperpforpoynting})
indicate that 
\begin{equation}
j_{P \perp}  \sim {{1}\over{b^2 f_{\perp}(b)}}
\label{jperpgeneraldep} \end{equation}
where $f_{\perp}(b)$ is a slowly increasing function.  We can now describe
how the different components of the Lorentz force decline with
distance on different magnetic surfaces. The shape of these surfaces
matters, since the relation between the cylindrical distance $r$ and
the radial distance $b$ depends on it.  It is found that
\begin{equation}
j_{P \perp} B_P \sim {{1}\over{b^4 f_{\perp}(b)}}
\label{torquingforceversusdist}
\end{equation}
\begin{equation}
j_{P \perp} B_{\theta} \sim {{1}\over{r b^2 f_{\perp}(b)}}
\label{accelerativeforceversusdist}
\end{equation}
\begin{equation}
j_{\theta}  B_P \sim {{1}\over{r^3 b^2}}
\label{pinchingfrompolversusdist}
\end{equation}
\begin{equation}
j_{P \|} B_{\theta} \sim  {{1}\over{ r b^{2 \gamma} I_{\infty}(b) }}
\label{hoopstressversusdist}
\end{equation}

The component of the Lorentz force accelerating the fluid along field
lines $j_{P \perp} B_{\theta}$, diminishes most slowly with
distance. This is particularly true for kinetic winds, as expected,
since these transform all of their Poynting energy into kinetic
energy.

\clearpage

\clearpage

\clearpage

\clearpage 

\clearpage 

\clearpage

\clearpage 

\clearpage 

\clearpage


\begin{thebibliography}{}
\bibitem[Anderson et al.(2003)]{anderson} Anderson, J.M., Li Z.Y., Krasnopolsky, R., Blandford, R.D. 2003, astro-ph/0304127
\bibitem[Begelman \& Li(1994)]{begli94} Begelman, M. C. and Li, Z. Y. 1994,  \apj, 426, 269
\bibitem[Beskin \& Okamoto(2000)]{beskinokamoto2000} Beskin, V.S.  and  Okamoto, I. 2000, \mnras, 313, 445
\bibitem[Blandford \& Payne(1982)]{blandfordpayne} Blandford, R. D. and Payne, D. G., 1982  \mnras, 199, 88
\bibitem[Bogovalov \& Tsinganos(1999)]{bogovalovtsin99} Bogovalov, S. V. and Tsinganos, K. 1999 \mnras, 305, 211
\bibitem[Bogovalov(1996)] {bogovalov96} Bogovalov, S.V. 1996, \mnras, 280, 39
\bibitem[Bogovalov(2000)]{bogovalov00} Bogovalov, S. V. 2000 \aap, 002, 001
\bibitem[Cao \& Spruit(2002)]{cao} Cao, X. \& Spruit, H.C. 2002, \aap, 385, 289
\bibitem[Contopoulos \& Lovelace(1994)]{ContopoulosLovelace94} Contopoulos, J.
and Lovelace, R.V.E. 1994, \apj,  429, 139
\bibitem[Heyvaerts(1996)]{HeyvMiniato} Heyvaerts, J. 1996, Plasma Astrophysics, Astrophysics School VII San Miniato, Italy, eds. C.Chiuderi \& G. Einaudi, Lecture Notes in Physics, {\bf 468}, 31 (Springer-Verlag, Berlin)
\bibitem[Heyvaerts \& Norman(1989)]{HeyvNorman89} Heyvaerts, J. and Norman, C.A. 
1989, \apj,  347, 1055
\bibitem[Heyvaerts \& Norman(2002a)]{HNII}
Heyvaerts, J. and Norman, C.A. 2002a, \apj, submitted
\bibitem[Heyvaerts \& Norman(2002b)]{HNIII}
Heyvaerts, J. and Norman, C.A. 2002b, \apj, submitted
\bibitem[Kudoh et al.(2002)]{Kudohetal2002} Kudoh, T., Matsumoto, R. \& Shibata, K. 2002 \pasp, 54, 267
\bibitem[Lery \& Frank(2000)]{LeryFrank} Lery, T. and Frank, A. 2000, \apj, 533, 897
\bibitem[Li(1993)]{Li93} Li, Z.Y. 1993, PhD Thesis, University of Colorado, Boulder
\bibitem[Lynden-Bell(1996)]{LyndenBell96} Lynden-Bell, D. 1996, \mnras, 279, 389
\bibitem[Lovelace et al.(1991)]{Lovelaceetal91} Lovelace, R.V.E, Berk, H.L. and Contopoulos, J. 1991, \apj,  379, 696
\bibitem[Mestel(1999)]{mestelbook} Mestel, L. 1999, Stellar Magnetism Oxford: Clarendon
\bibitem[Okamoto(1975)]{okamoto75} Okamoto, I. 1975,  \mnras,  173, 357
\bibitem[Okamoto(1999)]{okamoto99} Okamoto, I. 1999,  \mnras,  307, 253
\bibitem[Okamoto(2000)]{okamoto00} Okamoto, I. 2000,  \mnras,  318,  250
\bibitem[Ostriker(1997)]{EOstriker97} Ostriker, E. C. 1997, \apj, 486, 291 
\bibitem[Romanova et al.(1997)]{Romanovaetal97} Romanova, M. M., Ustyugova, G. V., Koldoba, A. V., Chechetkin, V. M. and  Lovelace, R. V. E.  1997, \apj, 482, 708
\bibitem[Ouyed and Pudritz(1997a)]{ouyedpudritz97a} Ouyed, R. and Pudritz, R. 1997a, \apj, 482, 712
\bibitem[Ouyed \& Pudritz(1997b)]{ouyedpudritz97b} Ouyed, R. and Pudritz, R. 1997b, \apj, 484, 794
\bibitem[Ouyed \& Pudritz(1999)]{ouyedpudritz99} Ouyed, R. and Pudritz, R. 1999, \mnras, 309, 233
\bibitem[Sakurai(1985)]{sakurai85} Sakurai, T. 1985,  \aap,  152, 121
\bibitem[Sauty et al.(1992a)]{tsinsauty92a} Tsinganos, K. and Sauty, C. 1992, \aap, 255, 405
\bibitem[Sauty et al.(1992b)]{tsinsauty92b} Tsinganos, K. and Sauty, C. 1992, \aap, 257, 790
\bibitem[Sauty \& Tsinganos(1994)]{sautyetal94} Sauty, C. and Tsinganos, K. 1994, \aap, 287, 93
\bibitem[Sauty et al.(1999)]{sautyetal99} Sauty, C. Tsinganos, K. and Trussoni, E. 1999, \aap, 348, 327
\bibitem[Shafranov(1996)]{shafranov} Shafranov, V. D. l966, {\it Reviews of Plasma Physics} , 2, 103 (New York: Consultants Bureau)
\bibitem[Shu et al.(1995)]{Shuetal95} Shu, F., Najita, J., Ostriker, E. C. and  Shang, H.  1995, \apjl, 455, 155
\bibitem[Tomimatsu(1995)]{Tomimatsu95} Tomimatsu, A. 1995 \pasj,  46, 123
\bibitem[Tomisaka(2002)]{tomisaka} Tomisaka, K. 2002, \apj, 575, 306
\bibitem[Trussoni et al.(1997)]{Trussonietal97} Trussoni, E., Tsinganos, K. and Sauty, C. 1997, \aap,  325, 1099
\bibitem[Tsinganos \& Bogovalov(2000)]{tsinbogovalov00} Tsinganos, K. and Bogovalov, S. V. 2000, \aap 356, 989
\bibitem[Ustyugova et al.(1995)]{Ustyugovaetal95} Ustyugova G. V., Koldoba, A. V., Romanova, M. M., Chechetkin, V. M. and  Lovelace, R. V. E.  1995, \apjl, 439, 39
\bibitem[Ustyugova et al.(1999)]{Ustyugovaetal99} Ustyugova G. V., Koldoba, A. V., Romanova, M. M., Chechetkin, V. M. and  Lovelace, R. V. E.  1999, \apj, 516, 221
\bibitem[Ustyugova et al.(2000)]{Ustyugovaetal00} Ustyugova G. V., Lovelace, R. V. E., Romanova, M. M., Li, H. and Colgate, S.A.  2000, \apjl, 541, 21
\bibitem[Vlahakis \& Tsinganos(1998)]{VlahakisTsing98} Vlahakis, N. and Tsinganos, K. 1998, \mnras, 298, 777
\bibitem[Vlahakis \& Tsinganos(1999)]{VlahakisTsing99} Vlahakis, N. and Tsinganos, K. 1999 \mnras, 307, 279
\bibitem[Weber \& Davis(1967)]{weberdavis67} Weber, E. J. and Davis, L. Jr. 1967, \apj, 148, 217

\end{thebibliography}
\end{document}